\documentclass[3p,times]{article} 

\usepackage{url}
\usepackage{amssymb}
\usepackage[utf8]{inputenc}
\usepackage{listings}
\usepackage{capt-of}
\usepackage{wrapfig}
\usepackage{graphicx}
\usepackage{subfigure}
\usepackage{caption}
\usepackage{geometry}
\usepackage{color}
\usepackage{xspace}

\newcommand{\doublefig}[8]{%
 \begin{figure*}[htp]%
  \centering%
  \subfigure[#1]{\includegraphics[width=.49\columnwidth]{#2}\label{#3}}%
  \hfill%
  \subfigure[#4]{\includegraphics[width=.49\columnwidth]{#5}\label{#6}}%
  \caption{#7}\label{#8}%
 \end{figure*}%
}

\newcommand{\singlefig}[3]{%
 \begin{figure}[htp]%
  \centering%
  \includegraphics[width=.6\columnwidth]{#1}%
  \captionof{figure}{#2}%
  \label{#3}%
 \end{figure}%
}

\newcommand{\naive}{na\"{\i}ve\xspace}

\definecolor{red}{rgb}{1,0,0}
\definecolor{blue}{rgb}{0,0,1}
\definecolor{grey}{rgb}{0.5,0.5,0.5}
\definecolor{violet}{rgb}{0.5,0,0.5}
\definecolor{midblue}{rgb}{0,0,0.75}
\definecolor{darkred}{rgb}{0.5,0,0}
\definecolor{darkblue}{rgb}{0,0,0.5}
\definecolor{darkgreen}{rgb}{0,0.5,0}
\definecolor{orchid}{rgb}{0.45,0.51,0.9}
\definecolor{midgrey}{rgb}{0.5,0.5,0.5}

\usepackage[affil-it]{authblk}

\usepackage{geometry}

\DeclareCaptionFormat{listing}{\rule{\dimexpr\textwidth\relax}{0.4pt}\par\vskip1pt~#1#2#3}
\newcommand{\longtitle}{Revisiting Actor Programming in C++}

\begin{document}

\widowpenalty10000
\clubpenalty10000



\title{\longtitle}

\author[1]{Dominik Charousset\thanks{Corresponding author: dominik.charousset@haw-hamburg.de}}

\author[1]{Raphael Hiesgen}

\author[1]{Thomas C. Schmidt}

\affil[1]{\small{Hamburg University of Applied Sciences

Internet Technologies Group

Department Informatik, HAW Hamburg

Berliner Tor 7, D-20099 Hamburg, Germany}}

\maketitle


\begin{abstract}
The actor model of computation has gained significant popularity over the last decade. Its high level of abstraction  makes it appealing for concurrent applications in parallel and distributed systems. However, designing a real-world actor framework that subsumes full scalability, strong reliability, and high resource efficiency requires  many conceptual and algorithmic additives to the original model.  

In this paper, we report on designing and building CAF, the ``C++ Actor Framework''.
CAF targets at providing a  concurrent and distributed native environment for scaling up to very large, high-performance applications, and equally well down to small constrained systems. We present the key specifications and design concepts---in particular a message-transparent architecture, type-safe message interfaces, and pattern matching facilities---that make native actors a viable approach for many robust, elastic, and highly distributed developments. We demonstrate the feasibility of CAF in three scenarios: first for elastic, upscaling environments, second for including heterogeneous hardware like GPGPUs, and third for distributed runtime systems.
Extensive performance evaluations indicate ideal runtime behaviour for up to 64 cores at very low memory footprint, or in the presence of GPUs.
In these tests, CAF continuously outperforms the competing actor environments Erlang, Charm++, SalsaLite, Scala, ActorFoundry, and even the OpenMPI.
\end{abstract}


\section{Introduction}
\label{sec:intro}

In recent times, an increasing number of applications requires very high performance for serving concurrent tasks. Hosted in elastic, virtualized environments, these programs often need to scale up instantaneously to satisfy high demands of many simultaneous users. Such use cases urge program developers to implement tasks concurrently wherever algorithmically feasible, so that running code can fully adapt to the varying resources of a cloud-type setting. However, dealing with concurrency is challenging and handwritten synchronisations easily lack performance, robustness, or both. 

At the low end, the emerging Internet of Things (IoT) pushes demand for applications that are widely distributed on a fine granular scale. Such loosely coupled, highly heterogeneous IoT environments require lightweight and robust application code which can quickly adapt to ever changing deployment conditions. Still, the majority of current applications in the IoT is built from low level primitives and lacks flexibility, portability, and reliability. The envisioned industrial-scale applications of the near future urge the need for an appropriate software paradigm that can be efficiently applied to the various deployment areas of the IoT.

Forty years ago, a seminal concept to the problems of concurrency and distribution has been formulated in the actor model by Hewitt, Bishop, and Steiger \cite{hbs-umafa-73}. With the introduction of a single primitive---called actor---for concurrent and distributed entities, the model separates the design of a software from its deployment at runtime. The high level of abstraction offered by this approach combined with its flexibility and efficiency makes it highly attractive for the parallel multicore systems of today, as well as for tasks distributed on Internet scale. As such, the actor concept is capable of providing answers to  urgent problems throughout the software industry and has been recognized as an important tool to make efficient use of the infrastructure.

On its long path from an early concept to a wide adoption in the real world, many contributions were needed in both conceptual modeling and practical realization.
In his seminal work, Agha \cite{a-amccd-86} introduced mailboxing for the message processing of actors and later laid out the fundament for open and dynamically reconfigurable actor systems \cite{amst-ttac-92}.
Actor-based languages like Erlang \cite{a-eslia-96} or SALSA Lite \cite{dv-slhar-14} and frameworks such as ActorFoundry---which is based on Kilim \cite{sm-kitaj-08}---have been released but remained in specific niches, or vendor-specific environments (e.g., Casablanca \cite{ms-c-12}).
Scala includes the actor-based framework Akka \cite{ti-a-12} as part of its standard distribution, after the actor model has proven attractive to application developers.
The application fields of the actor model also include cluster computing as demonstrated by the actor-inspired framework Charm++ \cite{kk-cppwm-96}.
In  previous work \cite{cshw-nassp-13,chs-ccafs-14}, we reported on our first steps for bringing a  C++ actor library to the native domain. 

In this work, we present the ``C++ Actor Framework'' (CAF)\footnote{http://www.actor-framework.org, a predecessor of CAF was named {\tt libcppa}}. CAF has evolved over the last four years to a full-fledged development platform---a domain-specific language in C++ and a powerful runtime environment. Moreover, CAF subsumes components for GPGPU computing, support for distributed actor debugging, and adaptations to a loose coupling for the IoT \cite{hcs-eatdp-14}. It has been adopted in several prominent application environments, among them scalable network forensics \cite{vcspw-nahsn-14}. 

In this article, we rethink the design of actors with CAF, donating special focus to the following  core contributions.

\begin{enumerate}
\item We introduce the scalable, message-transparent architecture of CAF along with core algorithms.
\item We enhance robustness of future actor programming by introducing type-safe message passing interfaces.
\item We illustrate the operations and use of the CAF pattern matching facility.
\item We lay out a scheduling infrastructure for the actor runtime environment that improves scaling  up to high numbers of concurrent processors.
\end{enumerate}
Thorough, comparative performance evaluations of the messaging core, the scalability at mixed tasks and commodity benchmarks, the memory consumption and its release, as well as GPGPU integration complete the presentation.

The remainder of this paper is organized as follows. 
In Section~\ref{sec:case_for_actors}, we re-position the use of actors and highlight our platform from a programming and performance perspective. The evolution of the actor approach up to current requirements are discussed in \S~\ref{sec:background} together with related work.  Section~\ref{sec:key_concepts} introduces key concepts and technologies that are central for the development of our actor framework.  The software design for type-safe messaging interfaces between actors are developed in Section~\ref{sec:types}, followed by the chapter on actor scheduling (\S\,\ref{sec:scheduling}). Extensive performance evaluations are shown in \S\,\ref{sec:eval}. We conclude in Section~\ref{sec:conclusion} and give an outlook on future research.

\section{The Case for Actors}
\label{sec:case_for_actors}

Prior to presenting details about the C++ Actor Framework, we want to shed light on this developing field from three motivating perspectives: characteristic use cases, typical programming, and performance. Starting with use cases, we highlight the following features.

\subsection{Characteristic Use Cases}

\paragraph{Elastic programming for adaptive deployment} The actor approach allows a general purpose software to safely scale up and down by orders of magnitudes, while running in local or distributed environments. Such extraordinary robustness is enabled by assigning small application tasks to a possibly very high number of actors at negligible cost. In consequence, such programs can dynamically adapt to their runtime environment while executing efficiently on a mobile, a many-core server, or Internet-wide distributed hosts. The default domain of this case lies naturally in the cloud.

\paragraph{End-to-end messaging at loose coupling} Actors directly exchange messages with each other, while the deployment specific message transport remains transparent. A hosting environment of the actor system may well admit loose coupling, using a stateless transactional transport for example. This extends traditional models of distributed programming like the Remote Method Invocation (RMI) \cite{w-rpcjr-98}, which enables direct access to remote methods but requires a tight coupling, or the REST \cite{ft-pdmwa-00} facade, which is designed for loose coupling but adds an indirection.  Typical applications can use this capability to directly exchange signals with remote controllers---a light bulb, for example---or to move software instances between sites without reconfiguration. It is worth noting that rigorously defined (typed) message interfaces allow for dedicated `firewall' control, and do not open new vulnerabilities to Internet hosts.

\paragraph{Seamless integration of heterogeneous hardware} The abstract transport binding of actors seamlessly covers dedicated peripheral channels connecting GPUs etc.,  semi-internal buses common to current heterogeneous multi-CPU architectures, ``big.LITTLE'' boards, as well as gateway-based network structures in current cars, buildings, or factories. This allows actor systems to transparently bridge architectural design gaps and frees software developers from crafting dedicated glue code. 

\subsection{Programming Paradigm}

We now want to highlight the actor-based abstraction mechanisms for programming as opposed to ``traditional'' object-oriented designs. Consider the class definition in listing~\ref{lst:kvp}.

\begin{lstlisting}[caption={Interface definition of a class-based Key-Value store.},label=lst:kvp]
class key_value_store {
 public:
  virtual ~key_value_store();
  virtual void put(key k, value v) = 0;
  virtual value get(key k) const = 0;
};
\end{lstlisting}

It defines an interface named \lstinline^key_value_store^ supporting put and get operations.
When accessible by multiple threads in parallel, the class has to be implemented in a thread-safe manner. A simple approach is to guard both member functions using a mutex in the implementing class. This approach does not scale well, mainly because readers block other readers.
More scalable approaches require a specific synchronization protocol that is based on read-write locks, for example.
Architecturally, though, lock-based solutions are best avoided, as locks do not compose \cite{sl-scr-05}. In any case, the interface itself is not aware of concurrency.

Listing \ref{lst:actor_kvp} illustrates the interface definition of an actor offering put and get operations in our framework. The interface type \lstinline^kvs_actor^ specifies a \emph{message passing interface}.

\begin{lstlisting}[caption={Interface definition of an actor-based Key-Value store.},label=lst:actor_kvp]
using kvs_actor = typed_actor<reacts_to<put_atom, key, value>,
                              replies_to<get_atom, key>::with<value>>;
\end{lstlisting}

Actors can be programed without knowledge about concurrency primitives, but at the same time support massively parallel access (cf. \S\,\ref{sec:eval:mailbox_performance}).
In the example, get requests are sequentially processed without further coordination, as there is no intra-actor concurrency.
For a further increase of parallelism, actors can explicitly redistribute tasks to a set of ``workers''.
Our message passing interface uses so-called atoms (cf. \S\, \ref{sec:atoms}) to identify specific operations instead of member function names.


\begin{lstlisting}[caption={Caller side of an actor interface using \lstinline^sync_send^.},label=lst:sync_send_kvs]
sync_send(kvs, get_atom::value, some_key).then(
  [=](value some_value) {
    cout << some_key << " => " << some_value << endl;
  }
);
\end{lstlisting}

\singlefig{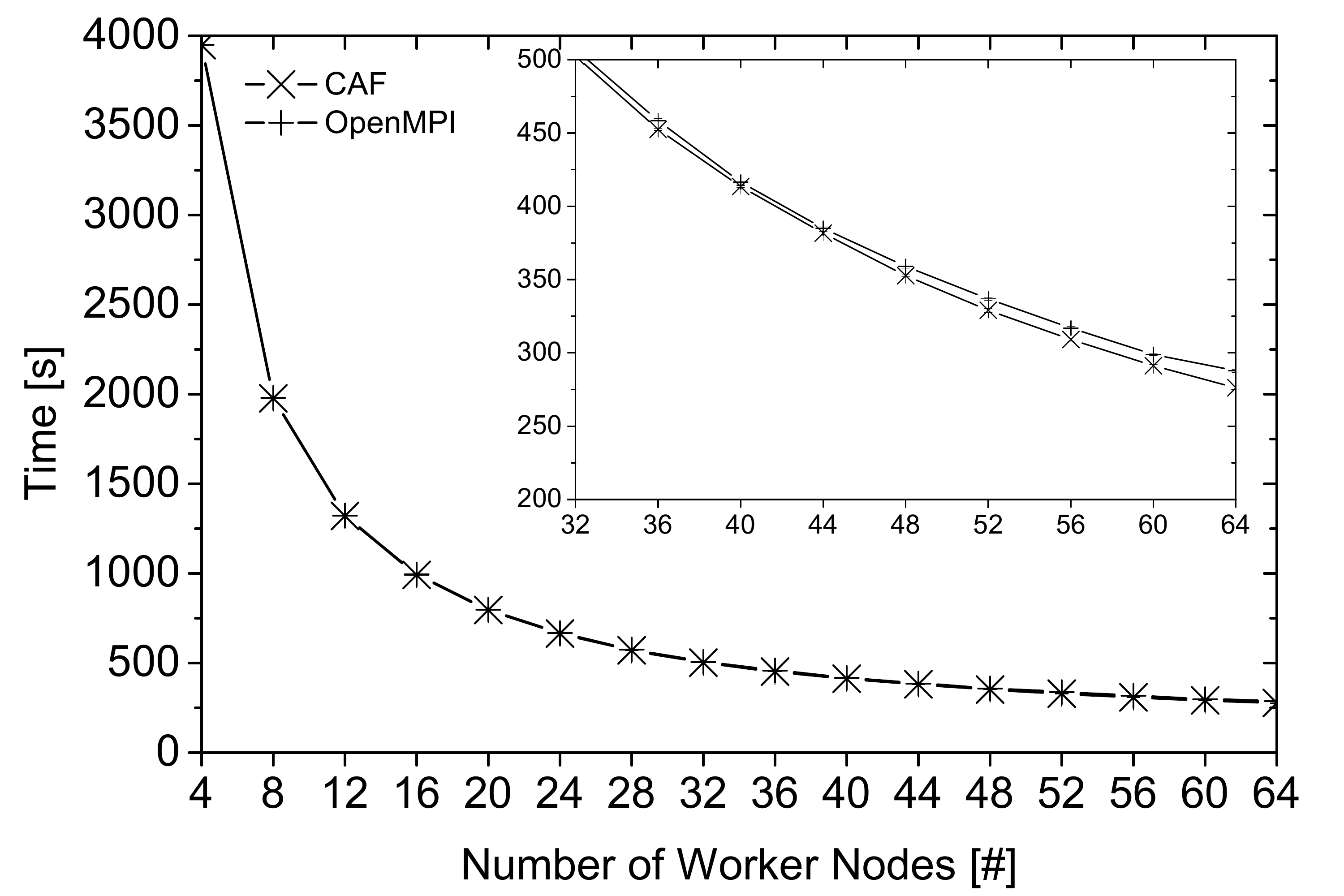}%
          {Sending and processing time for 3,000 images of the Mandelbrot set in a concurrent system consisting of 4-64 nodes using CAF and OpenMPI with magnification for 32--64 nodes}
          {fig:distributed_mandelbrot}

Our actor framework provides network-transparent messaging. When sending a message to a \lstinline^kvs_actor^---as shown in Listing \ref{lst:sync_send_kvs}---the sender can remain agnostic about where the receiver is located.
Caller and callee are also not coupled via the type system. This differs from RMI-based designs and reflects the desire for abstracting interfaces from implementations. 
The typed messaging handles enable the compiler to statically check input and output types, but do not expose the actual type of the callee.
The requester is rather able to send messages with partial type information, only, as the callee may implement additional message handlers for the \lstinline^kvs_actor^ interface not exposed to the caller.


\subsection{Competitive Performance}

Our third consideration is about performance. High-level abstractions in software design often disregard efficiency. We question whether appropriate performance can be achieved when switching from low-level message passing systems to actor frameworks. For an answer, we compare CAF with the well-known high-performance, low-level Message Passing Interface (MPI) \cite{sowdh-mcr-95}.   

We test  in a distributed system with an implementation to calculate a fixed number of the Mandelbrot set, using the same C++ program code with distribution done by CAF for actor programming in C++, and Boost.MPI (i.e., OpenMPI). 
Both versions exclusively rely on asynchronous communication and reduce synchronization steps to a minimum.
Since both programs share one C++ implementation for the calculation, the measurements reveal the overhead added by the distributed runtime system in use.
Hence, this setup discloses the trade-offs in performance which developers make when opting for a high-level abstraction like the actor model instead of low-layer primitives. Time measurements  were restricted to computations and do not include the initial setup phase performed by the platforms.

Figure \ref{fig:distributed_mandelbrot} shows the runtime results for concurrent setups as functions of available worker nodes.
In this evaluation, we have used one host machine running 4 to 64 virtual machines as worker nodes.
CAF runs 5\,\% faster with 64 worker nodes, indicating a slightly better scalability.
To achieve an evenly distributed work load, we added worker nodes in increments of four, as the host machine consists of four physical processors.

These results clearly illustrate that actors created from CAF do not impose a performance penalty when compared to a lower level message passing approach. Consequently, developers do not need to deal with a trade-off between a high level of abstraction on the one hand, and runtime performance on the other. An efficient implementation of the actor model can even outperform low-level approaches.

\section{The Evolution of Actors}
\label{sec:background}

Albeit formulated in the early 70s \cite{hbs-umafa-73},  until recently the actor model of computation mainly stayed with the Erlang community. The advent of multi-core machines and the growing importance of elastic cloud infrastructure made the model interesting for both academia and the software industry. In this section, we first discuss open conceptual questions in the current evolution of actor systems that arise from new application domains. Second, we contrast our approach with related work in the field of actor programming.

\subsection{Current Challenges}

The actor model allows to scale software from one to many cores and from one to many nodes. This flexibility in deployment makes the approach attractive for many application domains. This includes (1) infrastructure software, (2) Internet-wide distributed or IoT applications, as well as (3) high-performance applications that scale dynamically with demand. Still, available actor model implementations address only a subset of the requirements arising from these scenarios.

\paragraph{Robust Composition}

Initial definition and implementation of software is only a small part of its lifetime cycle. Maintenance and evolution are considered the main tasks of developers \cite{l-plcls-80}. Programming environments should support all parts of the lifetime cycle equally well and statically verify invariants. Still, available systems for actor programming either do not statically verify inter-relationships of actors, or require re-compilation and re-deployment even for minor changes in a monolithic code base. The first approach imposes excessive integration tests or constant model checking, while the latter is not suitable for large infrastructure software with independently maintained software components. Neither approach is robust with respect to existing compositions. Changing one part of the software in a downward compatible way must not affect other parts of the system, while invariants of a system should be statically verified at all times.

\paragraph{Native Programming}

We believe that writing dynamic, concurrent, and distributed applications using a native programming language such as C++ is ill-supported today. Standardized libraries only offer low-level primitives for concurrency such as locks and condition variables. It requires significant expert knowledge to  use such primitives correctly and can cause subtle errors that are hard to find \cite{ma-cpdcl-04}. A \naive memory layout may in addition severely slow down program execution due to \emph{false sharing} \cite{tlh-fsslm-94}. The support for distribution is even less advanced, and developers often fall back to hand-crafted networking components based on socket-layer communication. Transactional memory---supplied either in software \cite{st-stm-95} or hardware \cite{hm-tmasl-93}---and atomic operations can help implementing scalable data structures \cite{h-wfs-91}, but neither account for distribution, nor for communication between software components, nor for dynamic software deployment. A native programming environment based on the actor model gives developers full control over all performance-relevant aspects of a system while remaining at a very high level of abstraction that allows reasoning about the code without requiring expert knowledge on synchronization primitives.

\paragraph{Memory Efficiency}

Implementations of the actor model traditionally focus on virtualized environments such as the JVM \cite{ksa-afjpc-09}, while actor-inspired implementations for native programming languages focus on specific niches. For example, Charm++~\cite{kk-cppwm-96} is designed for software development for supercomputers. A general-purpose framework for actor programming that minimizes memory consumption is not available. Still, memory is the limiting resource on embedded devices and massively multi-user infrastructure software requires low per-user memory footprint at runtime in order to scale. In this way, a memory-efficient actor environment broadens the range of applications in both low-end and high-end computing.

\paragraph{Heterogeneous Hardware Environment}

Specialized hardware components are ubiquitously available. Modern graphics cards in commodity hardware are programmable via GPGPU languages such as CUDA \cite{nbgs-sppc-08} or OpenCL \cite{sgs-oppsh-10}. Algorithms running on such SIMD-components make use of hundreds or thousands of parallel processing units and outperform multi-core CPUs by orders of magnitudes for appropriate tasks. Custom hardware such as ASICs or reconfigurable hardware such as FPGAs achieve much higher performance than software for tasks like data encryption \cite{ch-rcsss-02}. Recent trends on mobile devices also couple two general-purpose processors of different complexity and speed on a single chip \cite{j-abtpe-12} that activates the fast but power consuming processor only when needed. In all cases, programmatic access to specialized hardware components requires to use a dedicated API. This makes integration of such devices into existing software systems cumbersome. A heterogeneous system architecture that transparently enables message passing between actors running on different hardware architectures helps developers to integrate heterogeneous components.

\subsection{Related Work}

The actor model was created by Hewitt, Bishop and Steiger \cite{hbs-umafa-73} and formalized by Agha et al. \cite{amst-ttac-92} to enable its use as a theoretical framework for modeling and verification languages such as Rebeca \cite{sj-tyaar-11}. The first de-facto implementation of the actor model with industrial applications was Erlang \cite{a-eslia-96}. While Hewitt et al. foresaw actors to monitor each other, Armstrong \cite{a-mrdsp-03} implemented a refined failure propagation model in Erlang to achieve reliability in the presence of hardware and software errors. This failure propagation model is based on monitoring, linking and hierarchical supervision trees and inspired most successive implementations, including CAF.

When multi-core processors became prevalent, intra-machine concurrency became relevant. Since thread-based solutions are inherently error-prone and non-composable \cite{l-tpwt-06}, developers started to seek better solutions. This lead to a growing interest in actor programming outside the Erlang community. As a result, it became important to provide actor implementations for other platforms. For this purpose, Haller and Odersky illustrated how event-based systems can be implemented without inversion of control \cite{ho-jmlc-06} to  build lightweight, event-based actor systems that are able to outperform thread-based approaches \cite{ho-sautb-09}. Lightweight actor systems like Akka \cite{ti-a-12} as well as CAF adopted parts of this implementation technique. Actor frameworks hosted by programming languages that allow sharing of state generally cannot ensure isolation of actors. At the same time, it gives developers access to a large set of well-tested special-purpose libraries that are unavailable in shielded actor frameworks \cite{tdj-wdsdm-13}. An exception to this is Kilim \cite{sm-kitaj-08} that ensures isolation of actors in Java applications using a bytecode postprocessor. With CAF, we explicitly decided to not use code generators or similar tools to require only a standard-compliant C++ compiler. In this way, we require no complex toolchain and are able to port CAF to many compilers and platforms.

Agha \cite{a-amccd-86} introduced mailbox-based message processing in his seminal modeling work on actors.
A mailbox is a FIFO ordered message buffer that is only readable by the actor owning it, while all other actors are allowed to enqueue  new messages. Mailboxes exclusively enable communication between actors, as no state is shared.  Implementation concepts of  mailbox management divide into two categories.
In the first category, an actor iterates over messages in its mailbox. On each receive call, it begins with the first but is free to skip messages.
As actors can change their behavior in response to a message,
a newly defined behavior may apply to previously skipped messages.
A message remains in the mailbox until it is eventually processed and removed as part of its consumption.
Erlang is the classical example for this category of message processing.

The second category of actor systems follows a more restrictive message processing scheme. A  message handler is invoked exactly once per message with the specific behavior of the actor. An untreated message cannot be recaptured at a later time, even though some systems allow to change the message handler at runtime. Consequently, actors are forced to treat messages in the order of arrival. The examples of Akka and Kilim fall in this category.

We follow the first approach, as it allows prioritizing messages and waiting for responses prior to returning to a default behavior. In this context, pattern matching has proven useful and very effective to ease definition of message handlers. Thus, we provide pattern matching for message handling as a domain-specific language (cf.~\S\,\ref{sec:types}). Further, we developed a concurrent queue algorithm tailored for use as mailbox (cf.~\S\,\ref{sec:lockfree-mailbox}).

A new concept introduced with CAF is its static verification of actor communication based on abstract messaging interfaces. The original actor model allows actors to send arbitrary data and requires the sender to dispatch on received content dynamically. This is natural in a dynamically typed language like Erlang, but also has been adopted in statically typed languages such as Scala or Java by the Akka framework for instance. Allowing arbitrary messages prevents the compiler from verifying the correctness of actor communication, i.e., validating that the behavior of the receiver includes a message handler for a given input. Not performing static checks for actor messages leaves correctness testing to the programmer, who is forced into excessive unit and integration tests. Static analysis tools such as Dialyzer~\cite{ls-dsdta-04} or dynamic model checkers such as McErlang~\cite{fs-mmcdf-07} can help finding bugs as long as the source code for all components is available (McErlang and other dynamic checkers require recompilation). Still, allowing arbitrary data as messages using dynamic typing techniques enables simplified access to actors with a common handle type for actors. This decouples callers from callees and allows programmers to modify an actor without updating or recompiling dependent actors.

Actor systems that statically type-check messages use an object-oriented design philosophy. Charm++ models actors as classes and represents actors with proxy instances on remote nodes. Invoking a member function implicitly uses message passing. The SALSA~\cite{va-pdros-01} programming language uses typed behaviors for actors that allows for static verification by the compiler. In both cases, callers are bound to the type of the callee. Changing the type of one actor requires a recompilation of all dependent actors, even for downward-compatible changes such as adding new handlers. In summary, the object-oriented design enables static type checking but introduces a tight coupling in the type system.

With CAF, we introduce a third approach that enables static verification of messaging without adding dependencies between caller and callee. Instead of exposing the type of an actor, we introduce abstract messaging interface definitions of pre-set semantics (cf.~\S\,\ref{sec:handles}). Actors can communicate even if only a subset of the messaging interface is known by the sender. Further, this enables actors to restrict available operations to others via the type system based on the context. In this way, we decouple callers and callees while preserving the static verification for actor messaging performed by the C++ compiler.

\section{The CAF Architecture and Key Concepts}
\label{sec:key_concepts}

The software design of CAF is based on a set of high-level goals, namely reliability, scalability, resource efficiency, and distribution transparency.  We want to make actor programming viable for a broad area of applications, ranging from (performance-) critical infrastructure software down to code running on embedded devices. All benefit from native execution and a low memory footprint, the latter being the limiting factor on embedded devices. A runtime that scales down to such constrained environments is required for bringing actor programming to the IoT.

From these high-level goals and  use cases, we can derive a number of key requirements. For reliability, type safety  is needed to provide a robust programming environment. Resource efficiency demands (1) to process messages efficiently for minimizing costs of the message-based abstraction, (2) a very low memory footprint of actors, and (3) a release of allocated memory as early as possible for re-use. Scalability foremost requires (1) to make efficient use of many CPU cores at minimal management overhead, (2) enable applications to use a large number of actors---thousands to millions---without performance penalty while consuming only a few hundred bytes per actor, and (3) span a distributed system over a few nodes up to hundreds of nodes while allowing dynamic rescale at runtime.

The design principles and algorithms that built CAF follow these goals and requirements as closely as possible. In addition, we adopt well-established design decisions that we consider best practice in actor systems such as the failure propagation model based on links and monitors known from Erlang for tightly coupled actors.

This section presents concepts and algorithms for building a programming environment that meets these criteria. We first give an overview of the software architecture of CAF in \S\,\ref{sec:architecture}. Important building blocks for the efficient messaging layer are presented in \S\,\ref{sec:lockfree-mailbox} and \S\,\ref{sec:copy-on-write}. Our software design enabling type-safe actor messaging interfaces is discussed in \S\,\ref{sec:atoms} and \S\,\ref{sec:handles}.
Lastly, \S\,\ref{sec:opencl-module} discusses our inclusion of GPGPU components.

\subsection{Architecture}
\label{sec:architecture}

Actors in CAF are hosted in a {\em runtime environment} that provides message dispatching, local scheduling, queue management, and everything required for adapting to the local deployment. Typically, a single runtime spans a full node in one distributed application. Actors communicate with other remote actors via this runtime environment in a transparent fashion. Even though only a single transport channel (port) needs to be exposed to the public, a multiplexer called {\em `middleman'} enables the mutually direct communication between thousands of actors on distinct machines. 

\begin{figure}[htp]
  \centering
  \includegraphics[width=.85\columnwidth]{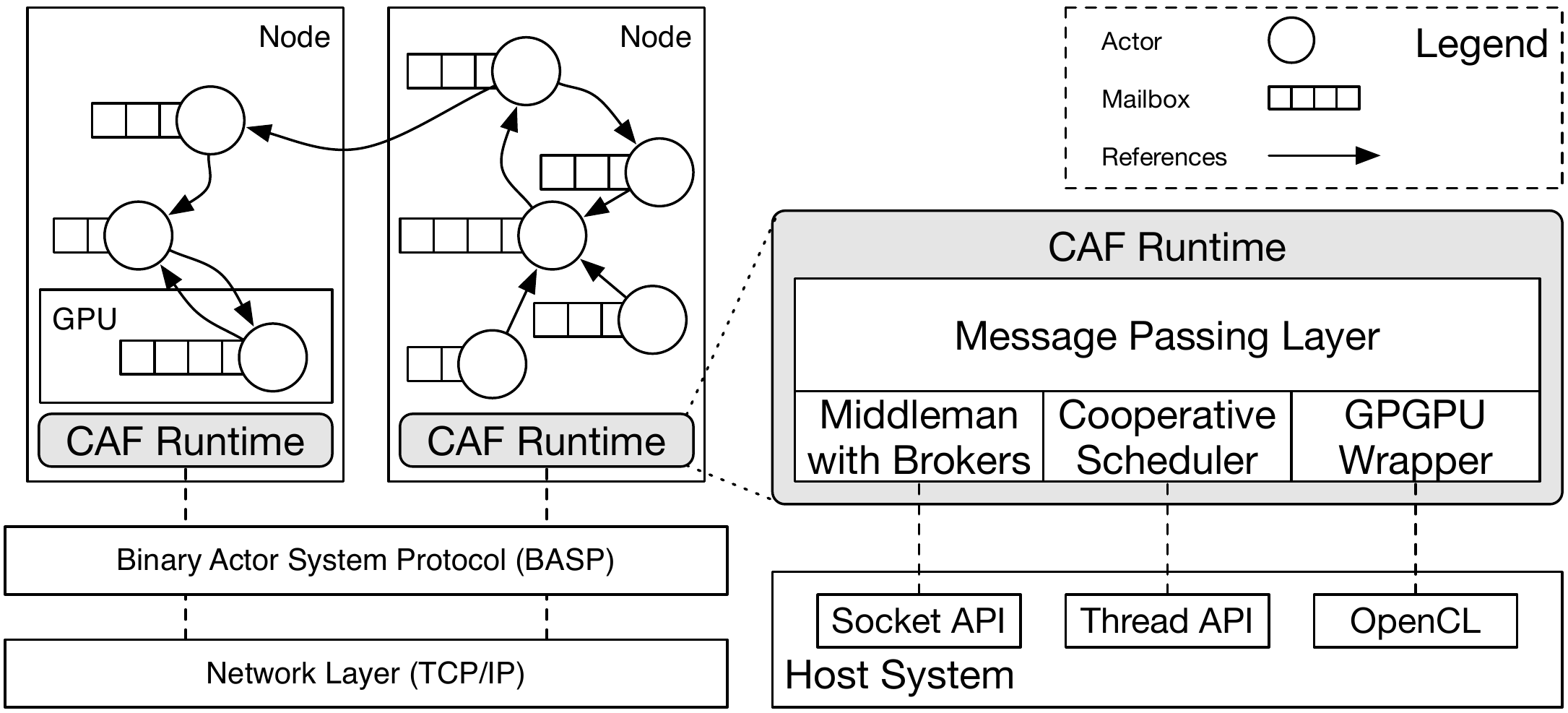}
  \captionof{figure}{The CAF architecture for a setup distributed across two nodes, running a few local actors each.}
  \label{fig:architecture}
\end{figure}

Figure \ref{fig:architecture} depicts the architecture of a  distributed CAF environment.  
 Actors see individual queues and are not aware of their physical deployment, but form a communication graph spanning multiple nodes. This flexible topology is enabled by the global message passing layer of the CAF runtime. The layer interconnects  components that implement individual services for actors, and multiple instances of the runtime exchange messages via the ``Binary Actor System Protocol'' (BASP).

Distributed runtime environments establish a common {\em message passing layer} via middlemen. The main function of a middleman is to organise the message exchange via networking interfaces like sockets. It multiplexes and encapsulates the network API of the host system to hide communication primitives. Packet and byte streams are converted to messages that are delivered to \emph{brokers}. A broker is an actor that performs asynchronous IO and lives in the event-loop of the middleman (cf.~\S\,\ref{sec:broker}). When an application starts, the CAF runtime instantiates an actor that implements BASP. The protocol transports actor messages and propagates errors from failing actors at the remote. Further, the ``BASP broker'' contacts remote actors on demand and transparently forwards inter-actor messages on the network.

The cooperative scheduler organises a  concurrent, fair execution of actors on a sub-thread level of the local host. It uses the threading API of the C++ standard library to manage  worker threads from the host system and distribute load among them. To perform the latter at runtime, the scheduler transparently dispatches message handlers from event-based actors to the workers. The design and implementation of the scheduler is discussed in detail in \S\,\ref{sec:scheduling}. It is one crucial component  of CAF---responsible for the concurrent system performance and its scalability.

Wrapper components may be added to hide heterogeneous hardware components behind a facade. Currently, the GPGPU wrapper  creates actors from OpenCL kernels as shown in \S\,\ref{sec:opencl-module}. At runtime, the wrapper transparently converts messages to OpenCL-compatible data and vice versa. Thereby, the wrapper enables a flexible integration of heterogeneous hardware components without requiring a manual setup.

\subsection{Lockfree Mailbox Algorithm}
\label{sec:lockfree-mailbox}

The message queue or \textit{mailbox} implementation is a critical component of any message passing system. All messages sent to an actor are delivered to its mailbox, which acts as a shared resource whenever an actor receives messages from multiple senders in parallel. Thus, the overall system~performance, foremost its scalability depends
significantly on the selected algorithm.

A mailbox is a single-reader-many-writer queue. It is exposed to parallel write access, but only the owning actor is allowed to dequeue a message. Hence, the dequeue operation does not need to support parallel access.

We achieved this by combining a lock-free stack implementation with a FIFO ordered queue as internal cache. A lock-free stack can be implemented using a single atomic compare-and-swap (CAS) operation. It does not suffer from the so called $ABA$ problem of concurrent access that can corrupt states in CAS-based systems \cite{ibmc-iseap-83} as the enqueue operation only needs to manipulate the \textit{tail} pointer. Without reordering, the dequeue operation would have to traverse the (LIFO-sorted) stack in order to find the oldest element.

\singlefig{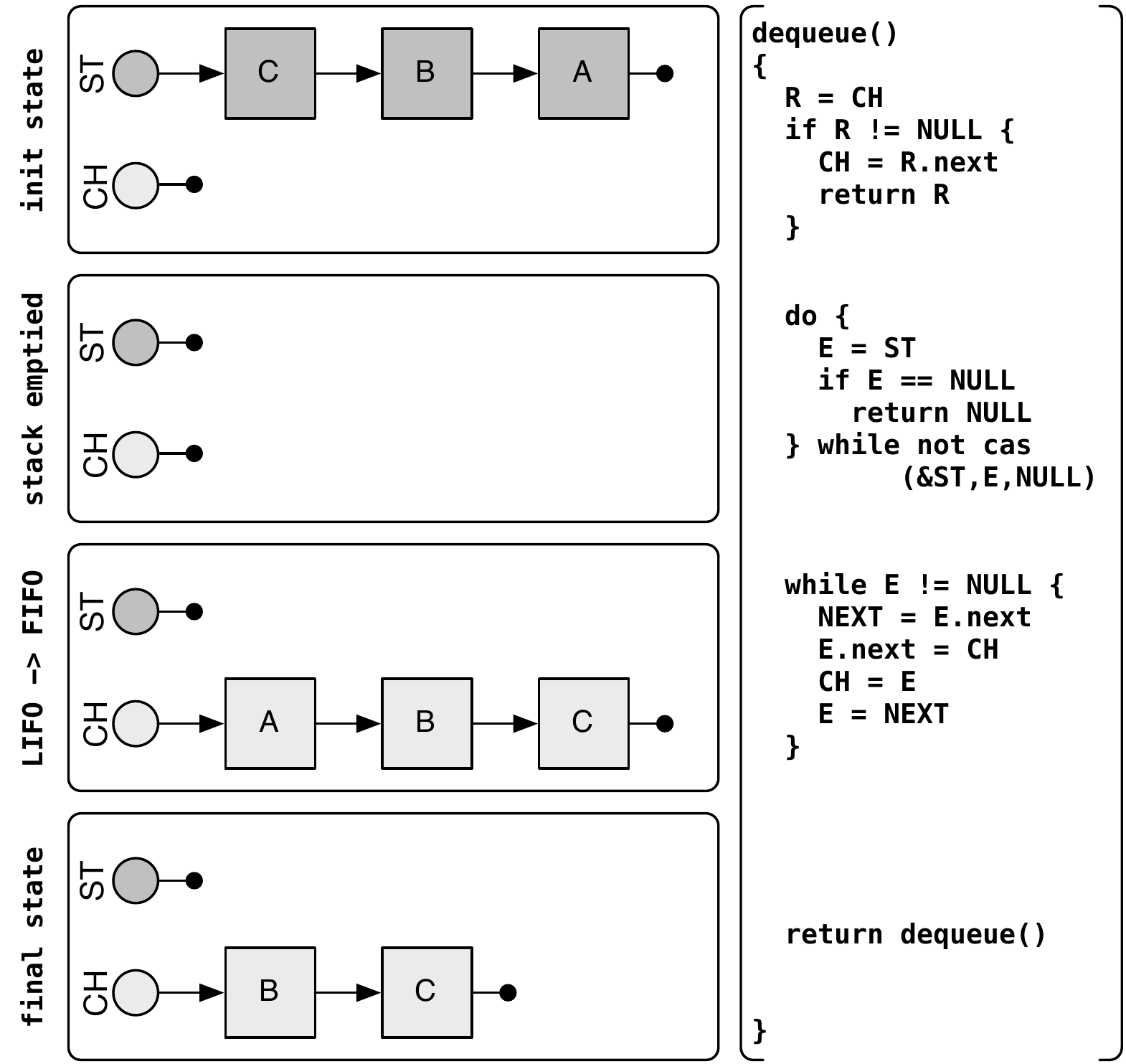}%
          {Dequeue operation in a cached stack (ST = Stack Tail, CH = Cache Head)}
          {fig:cached_stack_pop}

Figure \ref{fig:cached_stack_pop} shows the dequeue operation of our mailbox implementation. It always dequeues elements from the FIFO ordered cache (CH). The stack (ST) is emptied and its elements are moved in reverse order to the cache whenever it drains. Emptying the stack can be done by a single CAS operation as it only needs to set ST to NULL.

Our mailbox has complexity O(1) for enqueue operations, while the dequeue operation has an average runtime of O(1), but a worst case of O(n), where n is the maximum number of messages in the stack. Concurrent access to the cached stack is reduced to a minimum and both enqueueing and dequeueing perform only a single CAS operation. Our performance measurements (cf. \S\,\ref{sec:eval}) show that this lock-free implementation enables CAF to utilize hardware concurrency in N:1 communication scenarios more efficiently than common implementations of the actor model.

\subsection{Copy-on-write Messaging}
\label{sec:copy-on-write}

Copy-on-write is an optimization strategy to minimize copying overhead in a runtime instance of CAF. A message can be shared among several actors as long as all participants only demand read access. An actor implicitly copies the shared message when it requires write access and is only allowed to modify its own copy. Thus, data races cannot occur by design and each message is copied only if needed. This also implements garbage collection, as unreferenced messages are deleted automatically. As a result, message passing has call-by-value semantics from the perspective of a programmer. This eases reasoning about source code, since there is no complex lifetime management for messages.

We have used an atomic, intrusive reference counting smart pointer implementation that adds only a negligible runtime overhead. Any non-const dereferencing implicitly causes the smart pointer to detach its data whenever the reference count is greater than one. The overlaying pattern matching implementing is aware of this behavior and deduces const-ness from user-defined message handlers. In particular, the pattern matching engine will call the non-const dereference operator only if a message handler expects a mutable reference in its signature. In this way, CAF only relies on const correctness of user-generated code and does not impose any additional requirement on programmers to enable call-by-value semantics with implicit sharing. Pessimizations by the programmer---by taking arguments in a message handler by mutable reference when in fact no mutation takes place for example---can lead to unnecessary runtime overhead due to copying, but never affect the correctness of a program.

\subsection{Atom Constants: Type-safe Meta Information}
\label{sec:atoms}

Object-oriented designs use method names to unambiguously identify operations. When using a non class-based abstraction, programmers need an equivalently powerful way of encoding the target operation. Erlang introduced so-called \emph{atoms} for this purpose. An atom is a named constant that uniquely identifies an operation on the receiver. Since C++ does not support atoms natively, we contribute a design that for uniquely typed named constants with minimal runtime overhead.


\begin{lstlisting}[caption={Definition of atom values.},label=lst:atom]
atom_value x = atom("x");
auto hello = atom("Hello"); // automatically deduces `atom_value` for hello
\end{lstlisting}

Atoms are part of a message and serve as meta data. This makes an efficient processing of atoms mandatory, as they are always processed with the content of the message. Further, creating atoms must have negligible overhead, because senders frequently request atoms when sending data.

Listing \ref{lst:atom} illustrates the definition of two atoms in CAF named \lstinline^x^ and \lstinline^hello^. The function \lstinline^atom^ is declared \lstinline^constexpr^, meaning that it is evaluated at compile time. As a result, sending a message with atoms does in fact incur no runtime overhead. The compiler replaces the function call to \lstinline^atom^ with a constant 64-bit value generated from the given string literal. Thus, evaluating atoms at runtime has the overhead of one integer comparison.

In order to enforce static checking of messaging interfaces, we need to render the actual value of an atom visible to the compiler. Unfortunately, C++ cannot generate unique types for string literals, why the function \lstinline^atom^ always returns an \lstinline^atom_value^. To make these values accessible at compile time, we lift them to types by using the template class \lstinline^atom_constant^, as shown in Listing~\ref{lst:atom_constant}.

\begin{lstlisting}[caption={Definition of compile-time atom constants.},label=lst:atom_constant]
using x_atom = atom_constant<atom("x")>;
using hello_atom = atom_constant<atom("Hello")>;

auto value1 = x_atom::value; // value1 is of type x_atom
atom_value value2 = x_atom::value; // constants are convertible to atom_value
\end{lstlisting}

The template \lstinline^atom_constant^ is an implementation of the ``int-to-type'' idiom \cite{a-gpmtv-00} and enables programmers to use atoms in typed messaging interfaces. Each atom constant declares the static member \lstinline^value^ to get an instance of that particular type. This allows a seamless use of atoms as both type and value. Actors that match on \lstinline^atom_value^ will receive all atoms, while matching on a particular \lstinline^atom_constant^ will match exactly one value. In this regard, \lstinline^atom_value^ can be seen as common base type for all atom constants.

Our implementation of the function \lstinline^atom^ converts ASCII characters to a 6-bit encoding similar to Base64 \cite{rfc-4648}. This restricts the input string length to ten characters but provides a collision-free, reversible mapping. The remaining four bits are used as starting sequence to detect the position of the first character.

\subsection{Actor Handles}
\label{sec:handles}

Software entities in message passing systems have to characteristic attributes: identifiers and interfaces. The former address software entities---in our case actors--- in a network-transparent manner, while the latter encode valid inputs. For the remainder of this paper, we refer to messaging interfaces simply as \emph{interface} for brevity.

The original actor model uses \emph{mail addresses} as identifiers with an implicit wildcard interface \cite{hbs-umafa-73}. A wildcard interface accepts any input and the receiver is responsible to perform dynamic dispatching on received data, usually via pattern matching.

Most actor implementations closely follow this design. Either by using a dynamically typed programming language or by using type erasure techniques at the sender to allow arbitrary inputs. An example for such a design in a statically typed language is the \lstinline^ActorRef^ in Akka. This locator type is used directly for message passing and accepts inputs of type \lstinline^Any^ in Scala and \lstinline^Object^ in Java. This is the respective root of the class hierarchy in both languages. Such an approach hides information from the compiler, rendering a static analysis of the interfaces impossible.

Actor model implementations that do not use implicit wildcard interfaces such as SALSA or Charm++ use an object-oriented approach for defining interfaces that causes dependencies between senders and receivers and thus is not well-suited for open systems (cf. \S\,\ref{sec:background}). Interfaces in this approach are modeled as proxy objects that hide the identifier, but expose the type of the callee. Since proxy objects are bound to a specific type, interfaces providing the same operations are not interchangeable. Further, sub interfaces can only be emulated with complex and brittle inheritance hierarchies.

CAF contributes a new design that enables static type checking without introducing dependencies between senders and receivers. Our design discloses all type information to the compiler in order to enable globally type-checked messaging without relying on brittle OO-like inheritance hierarchies. By using a domain-specific interface definition, we further enable actors to send messages with only partial type information of the receiver. Also, our design explicitly distinguishes between identifiers and interfaces and makes both accessible to programmers. This explicit design is different from the original actor model and---to the best of our knowledge---unique to CAF.

An interface is a mapping from unique input types to output types in our system. Mappings are sets, i.e., the ordering of mapping rules in the source code does not matter and interfaces have subset semantics. Wherever an actor with interface $X$ is expected, programmers are allowed to pass an actor with interface $Y$ instead, as long as $X \subseteq Y$. This is also true across the network. When connecting to a remote actor, the expected interface must be passed as a parameter. The call succeeds if a connection could be established and the expected the interface is a subset of---or equal to---the published interface of an actor. The published, i.e., publicly available, interface can in turn be a subset of the full interface of an actor. This feature of CAF allows developers to add more handlers to any actor in the system without breaking existing compositions. In particular, a re-compilation of dependent actors is not necessary. The flexible subset semantics give programmers fine-grained control over accessibility of certain operations by hiding parts of an interface depending on the context.

A \emph{handle} in CAF stores the interface of an actor as well as the identifier. Further, the definition of an interface simultaneously specifies the handle type and are type aliases for the variadic template \lstinline^typed_actor<Ts...>^. The parameter pack \lstinline^Ts^ is a compile-time list of $input \rightarrow output$ rules. Each rule is specified using the notation \lstinline^replies_to<Xs...>::with<Ys...>^ or \lstinline^replies_to<Xs...>::with_either<Ys...>::or_else<Zs...>^. The latter allows programmers to specify operations that return \lstinline^Ys...^ on success and \lstinline^Zs...^ on failure. Operations that do not produce results can use \lstinline^reacts_to<Xs...>^ for convenience, which is an alias for \lstinline^replies_to<Xs...>::with<void>^. Listing \ref{lst:typed_actor_handles_examples} shows the definition of three different interfaces as type aliases, where \lstinline^adder^ is a subset of \lstinline^calculator^ and handles of the latter are consequently assignable to handles of type \lstinline^adder^.

\begin{lstlisting}[caption={Example declarations for typed actor handles.},label=lst:typed_actor_handles_examples]
using file_downloader =
  typed_actor<replies_to<url_atom, string>::with_either<ok_atom, string>
                                          ::or_else<error_atom>>;

using adder = typed_actor<replies_to<plus_atom, int, int>::with<int>>;

using calculator =
  typed_actor<replies_to<plus_atom, int, int>::with<int>,
              replies_to<minus_atom, int, int>::with<int>>;

void some_function() {
  calculator c = /*...*/; // get a calculator
  adder a = c; // assignment is allowed because adder is a subset of calculator
  // ...
}
\end{lstlisting}

CAF also does support explicit wildcard interfaces. This special case is modeled by the handle type \lstinline^actor^. Handles of this type are not assignable to typed handles and vice versa. Actors of this type are closer to the original actor model and can reduce code size when implementing a tightly coupled set of actors, e.g., when spawning local workers for single tasks. A more detailed discussion on typed vs. untyped handles can be found in \S\,\ref{sec:static_vs_dynamic}.

Both handle types can be queried to return the identifier of type \lstinline^actor_addr^. This identifier is used by the runtime to uniquely address and monitor actors in a distributed system. It can be used by programmers to determine whether two handles---possibly of different type---point to the same actor.

\subsection{Transparent Integration of GPGPU Hardware Components}
\label{sec:opencl-module}

With the advent of GPGPU programming, it became a crucial factor for a broad range of applications to make use of the heterogeneous computing platforms found in modern hardware deployments. This demand has lead to the development of the open standard OpenCL \cite{sgs-oppsh-10}. In OpenCL, developers provide an implementation of an algorithm, the so-called \emph{kernel}, in a C dialect that is compiled for the detected hardware at runtime. Listing~\ref{lst:opencl-kernel} shows the prototype of an OpenCL kernel to multiply two matrices.

\begin{lstlisting}[caption={Definition of an OpenCL kernel.},label=lst:opencl-kernel]
__kernel void matrix_multiply(
                __global float* matrix1,
                __global float* matrix2,
                __global float* output);
\end{lstlisting}

By convention, the last parameter is the output parameter. When instantiating this kernel at runtime to create an OpenCL \emph{program}, all three dimensions, i.e., the number of elements in \lstinline^matrix1^, \lstinline^matrix2^, and \lstinline^output^, must be defined. In order to execute \lstinline^matrix_multiply^, one needs to encapsulate the function call along with the parameters as a task and then enqueue this task to an OpenCL \emph{command queue}. OpenCL offers a callback-based API as well as a blocking API to await the completion of a task.

The task-based workflow of OpenCL is a natural fit to the actor model. Naturally, an OpenCL program can be regarded as an actor. It awaits input parameters and then produces results. In this exact way, CAF creates a message passing interface for OpenCL programs, as shown in the following example.

\begin{lstlisting}[caption={Creating a CAF actor from an OpenCL kernel.},label=lst:spawncl]
spawn_cl<float*(float*,float*)>(source, "matrix_multiply", {size, size});
\end{lstlisting}

The function \lstinline^spawn_cl^ expects the signature of the OpenCL kernel as a template parameter, normalized to a form with a result type instead of an implicit output parameter. The argument \lstinline^source^ is a string containing the source code of the kernel. The second argument is the name of the kernel. Finally, \lstinline^spawn_cl^ expects the dimensions of the input parameters. The invocation example shown above creates an actor that receives two arrays, each consisting of \lstinline^size^\,$\cdot$\,\lstinline^size^ (dimension on the x-axis multiplied with the dimension on the y-axis) elements, and replies a new array containing the resulting matrix. The matrices are represented in one dimension, since OpenCL does not support multi-dimensional arrays. The function \lstinline^spawn_cl^ also provides several overloads for fine-tuning the OpenCL behavior, or to perform data transformation. The latter allows to hide the kernel signature by providing a different interface to other actors. This is particularly useful to integrate OpenCL actors into an existing application.

\section{Defining Patterns and Actors in CAF}
\label{sec:types}

An actor is defined in terms of the messages it receives and sends.
Its behavior is hence specified as a set of message handlers that dispatch extracted data to associated functions.
Defining such handlers is a common and recurring task in actor programming.
The pattern matching facilities known from functional programming languages have proven to be a powerful, convenient and expressive way to define such message handlers.
Despite being recognized by the C++ community as a powerful abstraction mechanism \cite{sds-opmc-13}, there is neither language support nor a standardized API available yet.
However, pattern matching is a key ingredient for defining actors in a convenient and natural way.
Hence, we provide an internal domain-specific language (DSL) for this purpose.

\subsection{Pattern Matching Implementation}

Our DSL is limited to actor messages to keep the interface lightweight and focused on defining message handlers.
Unlike other runtime dispatching mechanisms, our pattern matching implementation discloses all types of incoming messages as well as the type of outgoing messages to the compiler.
In this way, the compiler can derive the interface of an actor from the definition of its behavior to perform static type checking.

A pattern in CAF is a list of \emph{match cases}. Each case is either (1) trivial, (2) a catch-all rule, or (3) an advanced expression enabling guards and projections. A trivial match case is generated from callbacks, usually lambda expressions. The input and output types are simply derived from the signature of the callback. A catch-all rule starts with \lstinline^others >>^, followed by a callback with zero arguments returning \lstinline^void^. A case of this kind always matches and produces no response message. An advanced match case begins with a call to the function \lstinline^on^ that returns an intermediate object providing the operator \lstinline^>>^. The right-hand side of the operator denotes a callback which should be invoked after a message matches the types derived from \lstinline^on^. Each argument to \lstinline^on^ is either a function object of signature \lstinline^T -> optional<U>^ or a value. The latter are automatically converted to function objects using a semantically equal function to \lstinline^to_guard^ shown in Listing \ref{lst:to_guard}. 

\begin{lstlisting}[caption={Lifting function for declaring guards from values.},label=lst:to_guard]
template <class T>
std::function<optional<T> (const T&)> to_guard(const T& value) {
  return [=](const T& other) -> optional<T> {
    if (value == other) {
      return value;
    }
    return none;
  };
}
\end{lstlisting}

We call function objects that map a value either to itself or to \lstinline^none^ \emph{guards}. They restrict the invocation of a callback based on the input value and forward the value itself to the callback. We further call functions that change the representation of a value \emph{projections}. An example for a projection is a string parser that tries to convert its input to an integer.

\begin{lstlisting}[caption={Usage examples for the pattern matching DSL of CAF.},label=lst:pattern_matching_examples]
auto odd_val = [](int i) -> optional<int> {
  if (i % 2 == 1) {
    return i;
  }
  return none;
};
auto str_float = [](const string& str) -> optional<float> {
  auto first = str.c_str();
  char* last;
  auto res = strtof(first, &last);
  if (last == first + str.size()) {
    return res;
  }
  return none;
};
behavior bhvr{
  on(42) >> [](int i) {
    assert(i == 42);
  },
  on(odd_val) >> [](int i) {
    // guaranteed to be odd
    assert(i % 2 == 1);
  },
  [](int i) {
    // only called if previous cases were *not* called
    assert(i % 2 == 0 && i != 42);
  },
  on(str_float) >> [](float f) {
    // input string could be converted to a float
  },
  [](const string& str) {
    // str could not be converted to a float
  }
};
\end{lstlisting}

Listing~\ref{lst:pattern_matching_examples} shows a mixed example using both trivial and advanced match cases. Line 1 declares a guard named \lstinline^odd_val^ that filters even integer values. Line 7 declares a projection that converts strings to floating point numbers using the C function \lstinline^strtof^. In line 16, we declare a local variable of type \lstinline^behavior^ and initialize it using a list of match cases. The first case in line 17 uses the convenience functionality of CAF to generate guards from values. The following lambda expression is called if and only if the received message consisted of the single integer value 42. The second match case in line 20 uses the \lstinline^odd_val^ guard and its associated lambda expression is only called for messages containing a single odd integer. The third (trivial) match case in line 24 is called whenever a message consisting of a single integer was received that was not matched by the previous two cases. Hence, the argument \lstinline^i^ inside the lambda expression can neither be odd nor 42. The fourth match case in line 28 uses the projection \lstinline^str_float^. It matches on messages consisting of a single string while the associated callback takes a \lstinline^float^. Whenever the conversion fails, the last case in line 31 gets called with the unchanged string.

Our DSL-based approach has more syntactic noise than a native support within the programming languages itself, for instance when compared to functional programming languages such as Haskell or Erlang.
However, we only use ISO C++ facilities, do not rely on brittle macro definitions, and our approach only adds negligible---if any---runtime overhead by making use of expression templates \cite{v-et-95}.
There is no additional compilation step required for the pattern matching.
Further, CAF does neither rely on code generators nor any vendor-specific compiler extension.

An important characteristic of our pattern matching engine is its tight coupling with the message passing layer.
The runtime system of CAF will create a response message from the value returned by the callback unless it returns \lstinline^void^.
Not only is this convenient for programmers, it also exposes the type of the response message to the type system.
This information is crucial to define type-safe messaging interfaces.

\subsection{Statically vs. Dynamically Typed Actors}
\label{sec:static_vs_dynamic}

CAF supports dynamically and statically typed actors. In both cases, programmers can either use free functions or classes.
All examples shown in the remainder of this section assume the definitions from Listing \ref{lst:boilerplate}.

\begin{lstlisting}[caption={Atom constants and type aliases used by subsequent listings.},label=lst:boilerplate]
using plus_atom = atom_constant<atom("plus")>;
using minus_atom = atom_constant<atom("minus")>;
using result_atom = atom_constant<atom("result")>;

using math_actor =
  typed_actor<replies_to<plus_atom, int, int>::with<result_atom, int>,
              replies_to<minus_atom, int, int>::with<result_atom, int>>;
\end{lstlisting}

A dynamic approach has the benefit of being able to provide \lstinline^actor^ a single handle type for all acquaintances.
This resembles the original actor modeling of Hewitt et al. that does not specify how---or even if---actors specify the interface for incoming and outgoing messages.
Rather, actors are defined in terms of \emph{names} they use, \emph{access rights} they grant to \emph{acquaintances}, and \emph{patterns} they specify to dispatch on the content of incoming data \cite{hbs-umafa-73}.

With statically typed actors, the compiler is able to verify the protocols between actors.
Hence, the compiler is able to rule out a whole category of runtime errors, because protocol violation cannot occur once the program has been compiled.
Note, that the compiler does not only verify the correct sending of a message but also the handling of the result when using \lstinline^sync_send^.
For instance, the example shown in Listing \ref{lst:type_mismatch} would be rejected by the compiler, because the client does expect the wrong type in the response message.

\begin{lstlisting}[caption={Type mismatch on the client side of a typed interface.},label=lst:type_mismatch]
math_actor ma = typed_spawn(typed_math);
sync_send(ma, add_atom, 10, 20).then(
  [](float result) {
    // compiler error: math actor will send an int as result, not a float
  }
);
\end{lstlisting}

When using \lstinline^sync_send^, a unique ID is assigned to the message.
The sending actor can use \lstinline^.then^ to install a message handler that is only used for the response to that particular ID.
The sender synchronizes with the receiver by skipping any other incoming message until it has either received the response message or an (optional) timeout has occurred.
Any error, e.g., if the sender no longer exists or is no longer reachable, will cause the sender to exit with non-normal exit reason unless it provides a custom error handler.

It is worth mentioning that the synchronization does not rely on blocking system calls and thus does not occupy any thread belonging to CAF.
Instead, any actor engaging in synchronous communication will simply not invoke any of its behavior-specific message handlers until the synchronous communication has taken place, ignoring all but the expected response message.

When using a statically typed system, developers are trading convenience for safety.
Since software systems grow with their lifetime and are exposed to many refactoring cycles, it is also likely that the interface of an actor is subject to change.
This is equivalent to the schema evolution problem in databases: once a single message type---either input or output---changes, developers need to locate and update all senders and receivers for that message.
When introducing a new kind of message to the system, developers also need to identify and update all possible receivers by hand.

With CAF, we lift the type system of C++ and make it applicable to the interfaces of actors.
At the same time, we are aware of the fact that dynamically typed systems do have their benefits and that these approaches are not mutually exclusive.
Rather, we believe a co-existence between the two empowers developers to make the ideal tradeoff between flexibility and safety.
Hence, we have implemented a hybrid system with CAF.
Type-safe and dynamic message passing interfaces are equally well supported and interaction between type-safe and dynamic actors is not restricted in any way.
From our experience, a good rule of thumb is that an actor should expose a typed interface whenever its visibility exceeds a single source file. In other words, actors with non-local dependencies should be checked by the compiler.
Such actors are usually central components of a larger system and offer a service to a set of actors that is either not known at coding time or might grow in the future.
Type-safe messaging interfaces allow the compiler to keep track of non-local dependencies that exist between central actors and a---possibly large---set of clients.

\subsection{Function-based Actors}
\label{sec:fun_based_actors}

Dynamically typed actors implemented as a free function can take an \lstinline^event_based_actor*^ as first argument, return a \lstinline^behavior^, or both. The first argument captures the implicit \lstinline^self^ pointer to the actor itself. A returned behavior is used to initialize the actor. The initial behavior can be replaced in response to a received message by calling \lstinline^self->become()^.

\begin{lstlisting}[caption={Example of a dynamically typed, function-based math actor.},label=lst:untyped_math_fun]
behavior math_fun(event_based_actor* self) {
  return {
    [](plus_atom, int a, int b) {
      return std::make_tuple(result_atom::value, a + b);
    },
    [](minus_atom, int a, int b) {
      return std::make_tuple(result_atom::value, a - b);
    },
    others >> [=] {
      cerr << "unexpected: " << to_string(self->current_message()) << endl;
    }
  };
}
\end{lstlisting}

Unhandled messages remain in the mailbox of an actor until it is eventually consumed. Whenever an actor receives a message that it does not handle in any state, this message remains indefinitely. To discard otherwise unmatched messages, an ``\lstinline^others^'' case can be used. A common work flow is to discard otherwise unhandled messages with an error report as shown in line 9--11 of Listing \ref{lst:untyped_math_fun}. It is worth mentioning that CAF follows the semantics of functional languages like Erlang or Haskell, i.e., the matching stops on the first hit. Any additional case after ``\lstinline^others^'' would be unreachable code.

Values returned from message handlers are automatically used as response to the sender. Returning multiple values can be achieved by returning a tuple as shown in line 4 and 7.

\begin{lstlisting}[caption={Example of a statically typed, function-based math actor.},label=lst:typed_math_fun]
math_actor::behavior_type typed_math_fun(math_actor::pointer self) {
  return {
    [](plus_atom, int a, int b) {
      return std::make_tuple(result_atom::value, a + b);
    },
    [](minus_atom, int a, int b) {
      return std::make_tuple(result_atom::value, a - b);
    }
  };
}
\end{lstlisting}

Listing \ref{lst:typed_math_fun} shows an equivalent actor implementing the interface \lstinline^math_actor^. The type for the \lstinline^self^ pointer can be obtained via \lstinline^math_actor::pointer^. The interface type also defines \lstinline^behavior_type^, which is a typed behavior allowing the compiler to statically verify the returned set of message handlers. Typed actors are allowed to change their state using \lstinline^self->become()^, which also expects a \lstinline^behavior_type^. Thus, actors must implement the proclaimed interface in each state. It is worth mentioning that using the \lstinline^others^ case from the previous example would result in a compiler error. Wildcards are not allowed in statically typed actors to enforce compliance to the implemented interface. Otherwise, subtle errors like typos or a change in the interface definition would cause unexpected behavior at runtime instead of being a compile-time error.

\subsection{Class-based Actors}
\label{sec:class_based_actors}

Dynamically typed actors implemented as a class derive from \lstinline^event_based_actor^ and must override the pure virtual member function \lstinline^make_behavior^.

\begin{lstlisting}[caption={Example of a dynamically typed, class-based math actor.},label=lst:untyped_math]
class math : public event_based_actor {
 public:
  behavior make_behavior() override {
    return {
      [](plus_atom, int a, int b) {
        return std::make_tuple(result_atom::value, a + b);
      },
      [](minus_atom, int a, int b) {
        return std::make_tuple(result_atom::value, a - b);
      },
      others >> [=] {
        cerr << "unexpected: " << to_string(current_message()) << endl;
      }
    };
  }
};
\end{lstlisting}

Listing \ref{lst:untyped_math} shows a class-based, dynamically typed actor with the same logic as implemented in \S\,\ref{sec:fun_based_actors}. Since the object itself is of type \lstinline^event_based_actor^, there is no need for capturing an explicit \lstinline^self^ pointer. Class-based actors are particularly useful for implementing actors with complex state, as managing data members can be more intuitive to programmers than recursively re-defining a behavior for state changes. There is no other benefit in using a class as opposed to using a free function.

\begin{lstlisting}[caption={Example of a statically typed, class-based math actor.},label=lst:typed_math]
class typed_math : public math_actor::base {
 public:
  behavior_type make_behavior() override {
    return {
      [](plus_atom, int a, int b) {
        return std::make_tuple(result_atom::value, a + b);
      },
      [](minus_atom, int a, int b) {
        return std::make_tuple(result_atom::value, a - b);
      }
    };
  }
};
\end{lstlisting}

Listing \ref{lst:typed_math} shows the implementation of an equivalent actor as seen in Listing \ref{lst:untyped_math} using the typed interface \lstinline^math_actor^. Each interface handle type defines \lstinline^base^ as an alias for \lstinline^typed_event_based_actor<...>^. This alias allows programmers to implement typed actors without repeating the interface. The type \lstinline^behavior_type^ is a typed behavior inherited from the base class.

\subsection{Brokers: Actors for Asynchronous IO}
\label{sec:broker}

When communicating with other services in the network, handling data packets or byte streams manually is often inevitable. For this reason, CAF provides \emph{brokers} as an actor-based abstraction mechanism over networking primitives. This is comparable to existing abstractions in Erlang or Akka. A broker is an event-based actor running in the so-called middleman. The middleman (MM) is a software entity that multiplexes low-level (socket) IO and enables late binding to platform-specific communication primitives. It translates network-layer events and byte streams to CAF messages as shown in Listing \ref{lst:brokermsgs}.

\begin{lstlisting}[caption={Special-purpose messages for
brokers.},label=lst:brokermsgs]
struct new_connection_msg {
  accept_handle source;
  connection_handle handle;
};
struct new_data_msg {
  connection_handle handle;
  std::vector<char> buf;
};
struct connection_closed_msg {
  connection_handle handle;
};
struct acceptor_closed_msg {
  accept_handle handle;
};
\end{lstlisting}

Brokers operate on any number of opaque handle types. Handles of type \lstinline^accept_handle^ identify a connection endpoint others can connect to. A \lstinline^connection_handle^ identifies a point-to-point byte stream, e.g., a TCP connection. Whenever a new connection is established, the MM sends a \lstinline^new_connection_msg^ to the associated broker. Messages of this type contain the handle that accepted the connection (\lstinline^source^) and a handle to the new connection (\lstinline^handle^). Whenever new data arrives, the MM sends a \lstinline^new_data_msg^ to the associated broker containing the source of this event (\lstinline^handle^) and a buffer containing the received bytes (\lstinline^buf^).

\begin{lstlisting}[caption={Example of a function-based broker.},label=lst:broker_example]
behavior mirror(broker* self) {
  return {
    [=](const new_connection_msg& msg) {
      self->configure_read(msg.handle, receive_policy::at_most(1024));
    },
    [=](const new_data_msg& msg) {
      self->write(msg.handle, msg.buf.size(), msg.buf.data());
    },
    others >> [] {
      // ignore anything else
    }
  };
}
\end{lstlisting}

Listing \ref{lst:broker_example} shows a function-based broker that writes back all data it receives. Brokers can configure how many bytes the MM aggregates before it sends a \lstinline^new_data_msg^ by setting a receive policy using the member function \lstinline^configure_read^ as shown in line 4. A policy configures either \emph{at least}, \emph{exactly}, or \emph{at most} a certain number of bytes and remains active until it is replaced. Outgoing data is written to an implicitly allocated buffer using the member function \lstinline^write^ as shown in line 7. It is worth noting that the broker does not contain any technology-specific information. It is usually bound during creation, e.g., to TCP port 42 using \lstinline^io::spawn_io_server(mirror, 42)^.

Since brokers run in the middleman and share a single IO event loop, implementations should be careful to consume as little time as possible in message handlers. Any considerable amount of computational work should be outsourced to other event-based actors. It is worth mentioning that the runtime implicitly salvages and re-uses existing buffers. Brokers run in the context of the MM. This means that the MM directly calls the message handler of its brokers with a message containing a \lstinline^new_data_msg^ for instance. The MM always uses the same message, re-writing its buffer over and over again. As long as the broker does not add a new reference count to this message, no copy of this buffer will ever be produced due to the copy-on-write optimization.

\section{Cooperative Scheduling Infrastructure}
\label{sec:scheduling}

The runtime of CAF maps N actors to M threads on the local machine. The number of threads depends on the number of available cores at runtime, while the number of actors dynamically grows and shrinks over the lifetime of an application. Actor-based applications scale by decomposing tasks into many independent steps that are spawned as actors. In this way, sequential computations performed by individual actors are small compared to the total runtime of the application, and the attainable speedup on multi-core hardware is maximized in agreement with Amdahl's law \cite{a-vspaa-67}.

Decomposing tasks implies that actors are often short-lived, why assigning a dedicated thread to each actor would not scale well. Instead, the runtime of CAF includes a scheduler that dynamically assigns actors to a pre-dimensioned set of worker threads. Actors are modeled as lightweight state machines that either 1) have at least one message in their mailbox and are \emph{ready}, 2) have no message and are \emph{waiting}, 3) are currently \emph{running}, or 4) have finished execution and are \emph{done}. Whenever a \emph{waiting} actor receives a message, it changes its state to \emph{ready} and is scheduled for execution. The runtime of CAF is implemented in user space and thus cannot interrupt running actors. As a result, actors that use blocking system calls such as IO functions can suspend threads and create an imbalance or lead to starvation. Such ``uncooperative'' actors can be explicitly detached by the programmer and assigned to a dedicated thread instead.

Here we focus on the scheduling of actors in a single runtime instance of CAF that is hosted on a single node. The remainder of this section presents the software design of our scheduling infrastructure in \S\,\ref{sec:scheduling-design} and discusses our deployed algorithm in \S\,\ref{sec:work-stealing}.

\subsection{Configurable, Policy-based Design}
\label{sec:scheduling-design}


The performance of actor-based applications depends on the scheduling algorithm in use and on its configuration. Different application scenarios require different trade-offs. For example, interactive applications such as shells or GUIs want to stay responsive to user input at all times, while batch processing applications demand to perform a given task in the shortest possible time. Programmers of the former applications want to minimize latency---the time between sending  a message to an actor and receiving its response, while programmers of the latter kind  only seek to maximize instructions performed per second. Actors operate on the granularity of individual messages and can be rescheduled at this pace. Allowing a running actor to drain its mailbox prior to rescheduling  maximizes the CPU time available to actors, and minimizes the efforts for the scheduler and for context switching. Actors with many messages in their mailbox, though, may delay execution of subsequent actors significantly, reduce agility, and increase the latency of the overall system.

Further, when running a CAF on a system that is shared with other demanding applications, developers may want to control the assignments of CPU cores to each application. Our design provides  default settings for general purpose use cases, but our API allows for configuring the dimensions  algorithm in use, maximum number of  messages processed per slot, and the number of worker threads.

\begin{lstlisting}[caption={Definition of a scheduler policy class.},label={lst:scheduler-policy}]
struct scheduler_policy {
  struct coordinator_data;
  struct worker_data;
  void central_enqueue(Coordinator* self, resumable* job);
  void external_enqueue(Worker* self, resumable* job);
  void internal_enqueue(Worker* self, resumable* job);
  void resume_job_later(Worker* self, resumable* job);
  resumable* dequeue(Worker* self);
  void before_resume(Worker* self, resumable* job);
  void after_resume(Worker* self, resumable* job);
  void after_completion(Worker* self, resumable* job);
};
\end{lstlisting}

Aside from managing the actors, the scheduler also bridges actor and non-actor code. For this reason, the scheduler distinguishes between external and internal events. An external event occurs whenever an actor is spawned from a non-actor context or an actor receives a message from a thread that is not under the control of the scheduler.

Listing~\ref{lst:scheduler-policy} shows the policy class to implement a scheduling algorithm.
Our scheduler consists of a single coordinator and a set of workers.
Note that the coordinator is needed by the public API to bridge actor and non-actor contexts, but is not necessarily an active software entity.
A policy provides the two data structures \lstinline^coordinator_data^ and \lstinline^worker_data^ that add additional data members to the coordinator and its workers respectively, e.g., work queues.
This grants developers full control over the full state of the scheduler.

Whenever a new work item is scheduled---usually by sending a message to an idle actor---, one of the functions \lstinline^central_enqueue^, \lstinline^external_enqueue^, and \lstinline^internal_enqueue^ is called.
The first function is called whenever non-actor code interacts with the actor system. For example when spawning an actor from \lstinline^main^.
Its first argument is a pointer to the coordinator singleton and the second argument is the new work item---usually an actor that became ready.
The function \lstinline^external_enqueue^ is never called directly by CAF. Its purpose is to model the transfer of a task to a worker by the coordinator or an other worker. Its first argument is the worker receiving the new task referenced in the second argument. 
The third function, \lstinline^internal_enqueue^, is called whenever an actor interacts with other actors in the system. Its first argument is the current worker and the second argument is the new work item.

Actors reaching the maximum number of messages per run are re-scheduled with \lstinline^resume_job_later^ and workers acquire new work by calling \lstinline^dequeue^. The two functions \lstinline^before_resume^ and \lstinline^after_resume^ allow programmers to measure individual actor runtime, while \lstinline^after_completion^ allows to execute custom code whenever a work item has finished execution by changing its state do \emph{done}, but before it is destroyed.
In this way, the last three functions enable developers to gain fine-grained insight into the scheduling order and individual execution times.

It is worth mentioning that \lstinline^scheduler_policy^ is not a base class. Rather, it is a concept class that shows all required types and function a policy must provide for the templated scheduler implementation.

\begin{lstlisting}[caption={Signature of \lstinline^set_scheduler^.},label={lst:set-scheduler}]
template <class Policy = default_scheduling_policy>
void set_scheduler(size_t num_workers = std::thread::hardware_concurrency(),
                   size_t max_msgs = std::numeric_limits<size_t>::max());
\end{lstlisting}

Listing~\ref{lst:set-scheduler} shows the prototype of \lstinline^set_scheduler^. This function allows programmers to configure all three parameters of the scheduler. The algorithm can be changed by setting the template parameter \lstinline^Policy^, which defaults to the default algorithm discussed in \S\,\ref{sec:work-stealing}. The first function argument configures the number of workers and the second argument the number of messages each actor is allowed to consume in a single run. The former defaults to the number of available CPU cores and the latter defaults to the maximum integer of value for \lstinline^size_t^, i.e., approximates no limit.

\subsection{Scheduling Algorithm}
\label{sec:work-stealing}

CAF is a general-purpose framework for actor programming. Hence, the default implementation of CAF should cover the majority of common use cases, while the selection of an appropriate algorithm is constrained in the following two dimensions.

The first thing to consider when choosing a scheduling algorithm is the architecture of multi-core machines. Independent of manufacturer, all multi-core designs---regardless of the number of processors on the die---use a single interconnect to coordinate memory access \cite{kzt-imaum-05}. Modifying the same memory region from multiple threads in parallel causes contention on the shared interconnect and severely slows down execution \cite{dhw-csma-97}. In this way, the hardware penalizes applications with frequent communication between threads and sets strict limits on the scalability of centralized scheduler designs.

The second consideration is about the scheduled entities. CAF is not limited to a particular application domain and thus can neither make assumptions on the average runtime of an actor nor on the spawn behavior. Instead, the scheduler of CAF is oblivious to the tasks it schedules and it cannot distribute tasks proactively, because CAF is unaware of the application logic.

The basic algorithm for oblivious scheduling in multiprogrammed environments is work stealing. The original algorithm was developed for fully strict computations by Blumofe et al. \cite{bl-smcws-94} and has an expected execution time of $T_{1}/P+T_{\infty}$ (which was also verified empirically), where $P$ is the number of workers, $T_1$ is the computation time on a single CPU and $T_{\infty}$ is the minimum execution time with an infinite number of processors. Communication overhead between workers is at most $P \cdot T_{\infty} \cdot (1 + n_d) \cdot S_{max}$, where $n_d$ is the maximum number of synchronizations of a single thread and $S_{max}$ is the largest activation record of any thread. Later extensions to the original design of the algorithm proved its applicability to arbitrary concurrent computations \cite{abp-tsmm-98}.

A work stealing scheduler has no shared work queue, but operates on individual queues of each worker. Workers dequeue work items only from their own queue until this is empty. Once this happens, the worker becomes a \emph{thief}, picking one of the other workers---usually at random---as a \emph{victim} and tries to \emph{steal} a work item from its queue. As a consequence, tasks (actors) are bound to workers by default and only migrate between threads as a result of stealing.

Examples that use work stealing include Cilk \cite{bjklr-cemrs-95}, Cilk++ \cite{l-ccp-09}, Java fork-join \cite{l-jff-00} (wich is used by Akka), X10 \cite{abbss-dsxcb-07}, Intel's Threading Building Blocks \cite{r-itbbo-07}, NUMA-aware OpenMP schedulers \cite{opwsp-otssm-12}, and parallelized implementations of the STL \cite{fs-pbosd-08}. Further, Lifflander et al. demonstrated that a hierarchical version of the algorithm scales to cluster deployments with up to 163,840 cores \cite{lkk-wspbi-12}.

Our implementation of the scheduler equips each worker with a double-ended, concurrent queue. Workers dequeue elements from the front of their own queue, but steal elements of other workers from the back. In this way, workers steal actors that suffered from longest waiting time. Top-level spawns---spawns without a parent---and messages from threads that do not belong to a worker of CAF add new elements to be back of the queue, while new items generated in the thread of a worker add new elements to the front. This approach increases cache locality, since the message causing an actor to be scheduled is likely still in the cache.

In relation to the policy interface described in \S\,\ref{sec:scheduling-design}, the default implementation is mapped as follows.
The \lstinline^coordinator_data^ contains only a single atomic integer, while \lstinline^worker_data^ contains the double-ended queue and a random-number generator. Calls to \lstinline^external_enqueue^ on the workers add elements to the back, while calls to \lstinline^internal_enqueue^ add elements to the front. Top-level spawns---a spawn without a parent---and messages from non-actors result in calls to \lstinline^central_enqueue^, which dispatches the item in round-robin order (using the atomic integer) to workers with \lstinline^external_enqueue^. The round robin order enables an even distribution during the bootstrapping phase, while the first actors are spawned (usually from \lstinline^main^). The function \lstinline^resume_job_later^ calls \lstinline^external_enqueue^ on the same worker.



\section{Performance Evaluation}
\label{sec:eval}

In this section, we analyze the performance of CAF.
The study focuses on the scalability of our software in comparison to other common actor systems and extends our previous work \cite{chs-ccafs-14}.
We want to examine the scaling behavior in terms of CPU utilization and memory efficiency.
Our host system is equipped with four 16-core AMD Opteron processors at 2299\,MHz each and runs Linux.
First, we perform two micro benchmarks on actor creation and mailbox efficiency, and second we run a larger scenario of mixed operations.
To include a benchmark created by the community, we adopt a Mandelbrot calculation from the Computer Language Benchmarks Game for our final benchmark.
The last benchmark we present examines the scalability when pushing parts of a workload to the GPU.

For comparative references, we use ActorFoundry, Charm++, Erlang, SALSA Lite, and Scala with the Akka toolkit.
In detail, our benchmarks are based on the following implementations of the actor model: (1) C++ with CAF 0.13.2 (\emph{CAF}) and Charm++ 6.5.1 (\emph{Charm}), (2) Java with ActorFoundry 1.0 (\emph{ActorFoundry}), (3) Erlang in version 5.10.2 using HiPE for native code generation and optimization level O3 (\emph{Erlang}), (4) the latest alpha release of the SALSA Lite programming language (\emph{SalsaLite}) and (5) Scala 2.10.3 with the Akka toolkit (\emph{Scala}).
CAF and Charm++ have been compiled as release versions using the Clang C++ compiler in version 3.5.2.
Scala, SALSA Lite and ActorFoundry run on a JVM configured with a maximum of 10\,GB of RAM.
For compiling ActorFoundry, we use the Java compiler in version 1.6.0\_38, since this version is required by its bytecode post-processor.

We measure both clock time and memory consumption.
Measurements are averaged over 10 independent runs to eliminate statistical fluctuations.
The memory consumption is recorded every 50\,ms during the runtime and the results are visualized as box plots to represent their variability in a transparent way.
Each box plot depicts a box containing 50\,\% of all measured values limited by the first quartile at its bottom and the third one at its top.
The median is shown as a band inside of the box and the mean is marked with a small square.
In addition, the whiskers mark the 5th and 95th percentiles, while the 1st and 99th percentiles are marked with crosses.
All graphs visualizing clock time are plotted with an error bar according to the 95\,\% confidence interval.
Our source code for all benchmark programs is published online at
\mbox{\url{github.com/actor-framework/benchmarks}}.

\subsection{Overhead of Actor Creation}

Our first benchmark reflects a simple divide \& conquer algorithm.
It computes $2^{20}$ by recursively creating actors.
In each step $N$, an actor spawns two additional actors of recursion counter $N-1$ and waits for the (sub) results of the recursive descend.
This benchmark creates more than one million actors, primarily revealing the overhead for actor creation.
Note that this algorithm does not imply the coexistence of one million actors at the same time.

\doublefig{Actor creation time}%
          {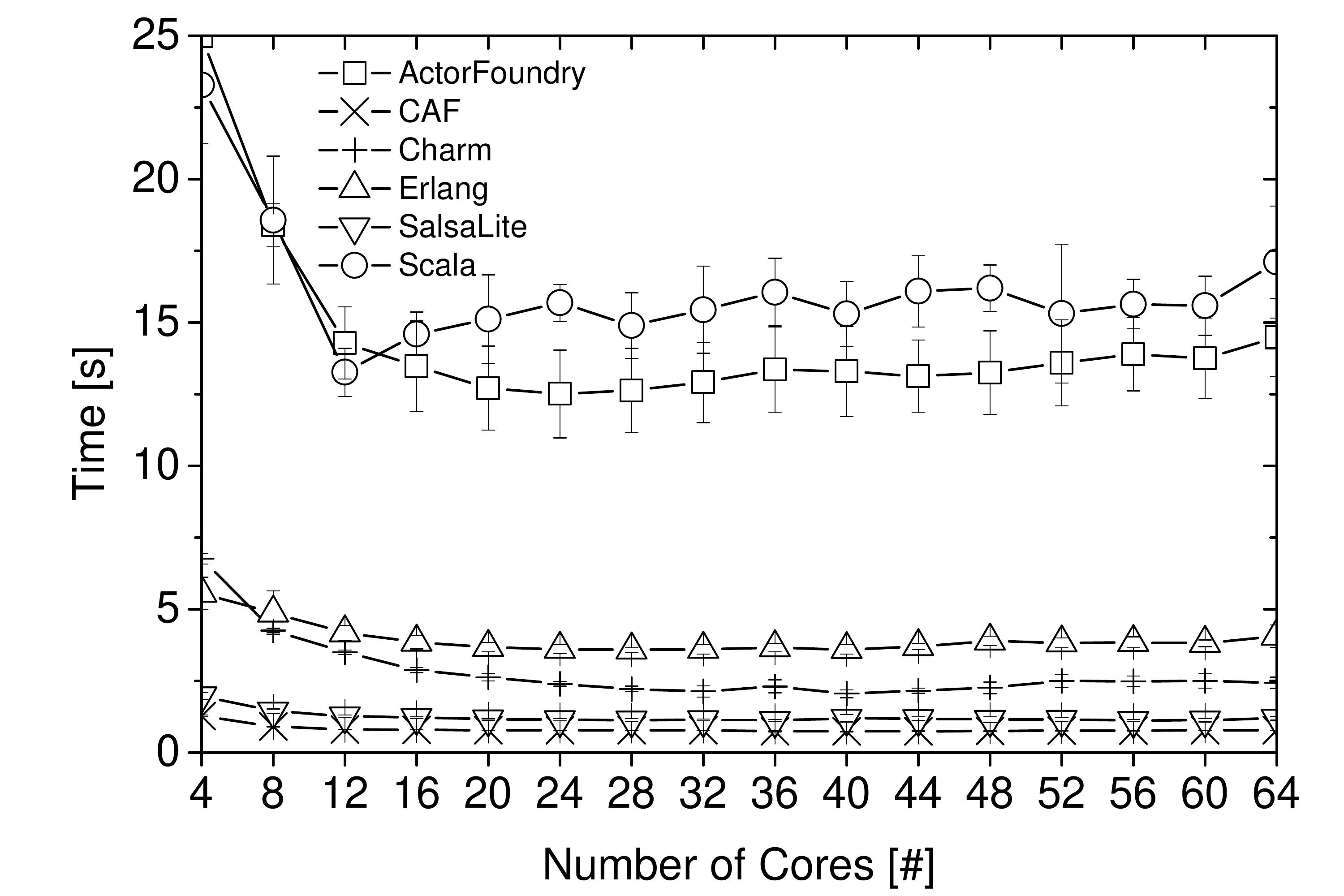}%
          {fig:creation:time}%
          {Memory consumption}%
          {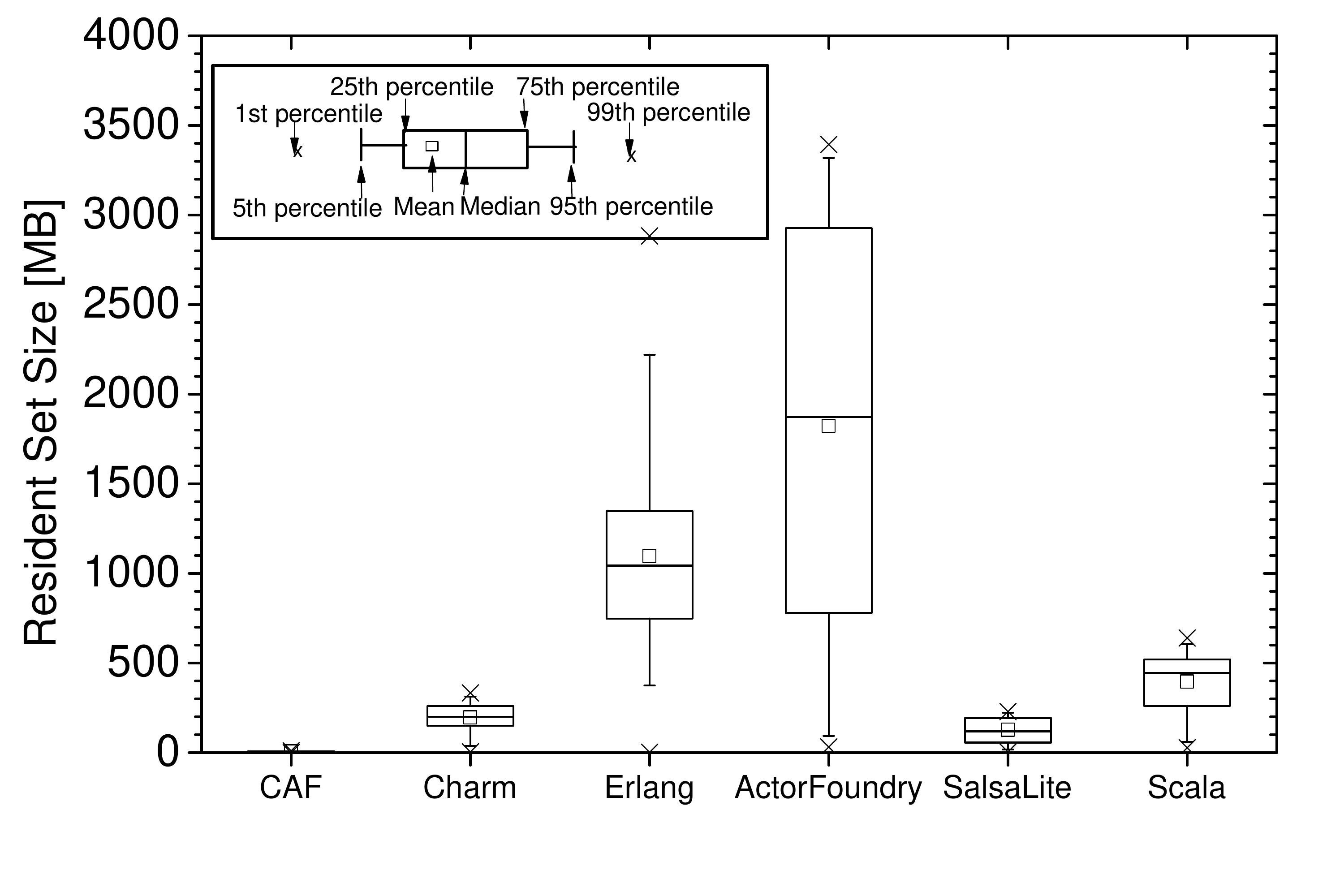}%
          {fig:creation:memory}%
          {Actor creation performance for $2^{20}$ actors}%
          {fig:creation}

Figure~\ref{fig:creation:time} displays the time consumed by this task as a function of available CPU cores.
CAF and SALSA Lite scale down nicely with cores, i.e., the scheduling of actor creation parallelizes cleanly for them.
While Charm++ does exhibit scaling behavior up to 32 cores, subsequent measurements are slightly higher in some cases.
In contrast, ActorFoundry only exhibits scalability up to 24 cores and Scala increases significantly in runtime after reaching a global minimum at 12 cores.
CAF is the fastest implementation with less than a second on eight or more cores.
On 64 cores, CAF, SALSA Lite and Charm++ run the benchmark in 3 or less seconds.
In contrast, Scala and ActorFoundry require 17 and 14 seconds respectively, while the high error bars indicate heavily fluctuating values.

Figure~\ref{fig:creation:memory} compares the memory consumption during this benchmark.
Results vary a lot in values and spread.
Notably, the highest values measured for CAF, Charm, SALSA Lite and Scala are lower than 75\% of all recorded values for Erlang and ActorFoundry.
ActorFoundry allocates significantly more memory than all other implementations, peaking around 3.5\,GB of RAM with an average of $\approx$\,1.8\,GB.
Erlang follows with a spike above 2\,GB of RAM and an average of $\approx$\,1\,GB.
Scala has an average RAM consumption of 500\,MB, with a spike at about 750\,MB.
SALSA Lite and Charm++ stay below 300\,MB, while CAF consumes about 10\,MB.
This low limit does not imply that an actor uses less than 10 Bytes in CAF.
CAF merely releases system resources as soon as possible and efficiently re-uses memory from completed actors.
A correlation of memory and runtime performance is not apparent.
Although Scala showed the worse performance, it does consume a medium amount of memory.

\subsection{Mailbox Performance in N:1 Communication Scenario}
\label{sec:eval:mailbox_performance}

Our second benchmark measures the performance in an N:1 communication scenario.
This communication pattern can be frequently observed in actor programs, typically whenever an actor distributes tasks by spawning a series of workers and awaits the results.

We use 100 actors, each sending 1,000,000 messages to a single receiver.
The minimal runtime of this benchmark is the time the receiving actor needs to process its 100,000,000 messages.
It is to be expected that the runtime increases with cores, because adding more hardware concurrency increases the speed of the senders and thus the probability of write conflicts.

\doublefig{Sending and processing time}%
          {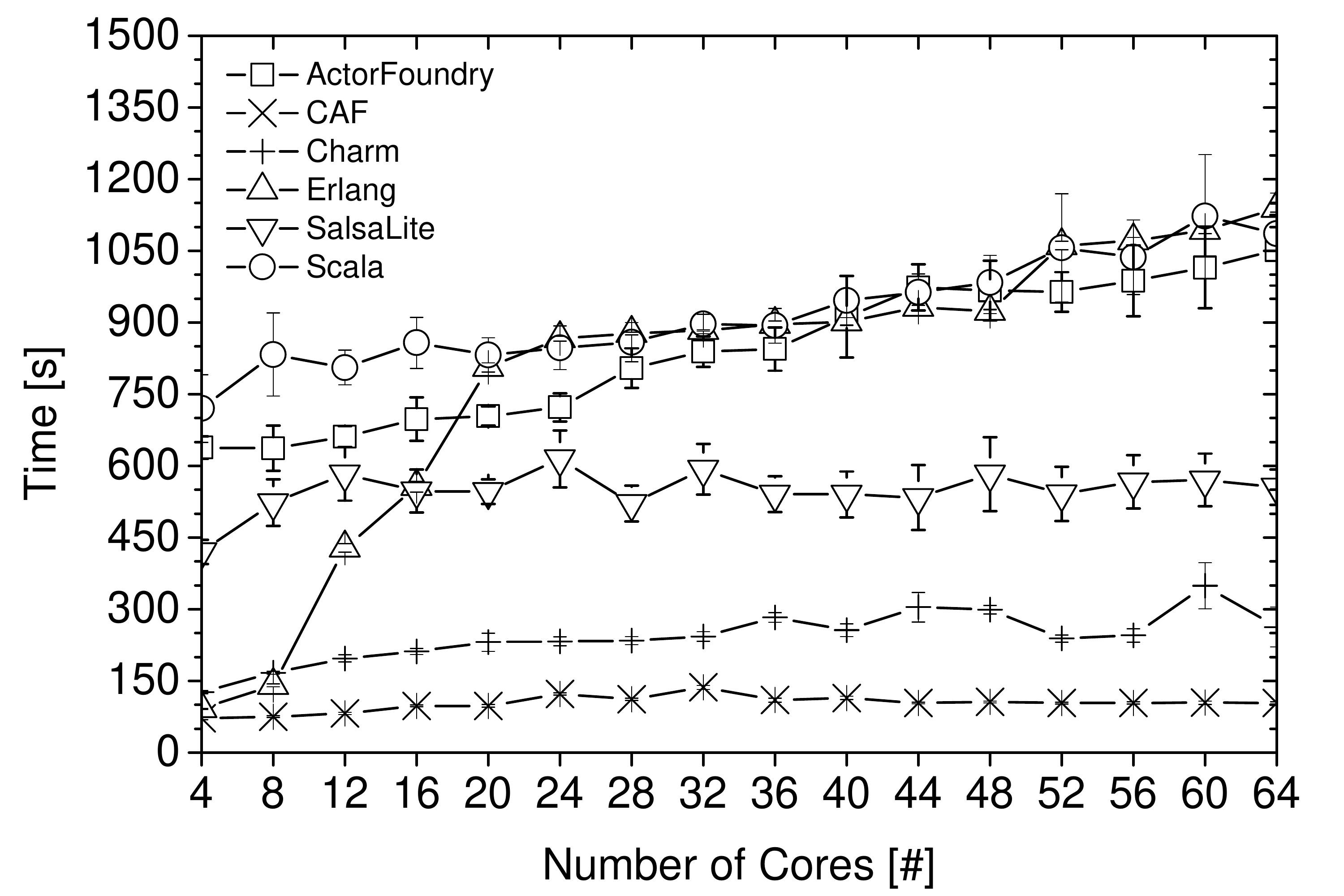}%
          {fig:mailbox:time}%
          {Memory consumption}%
          {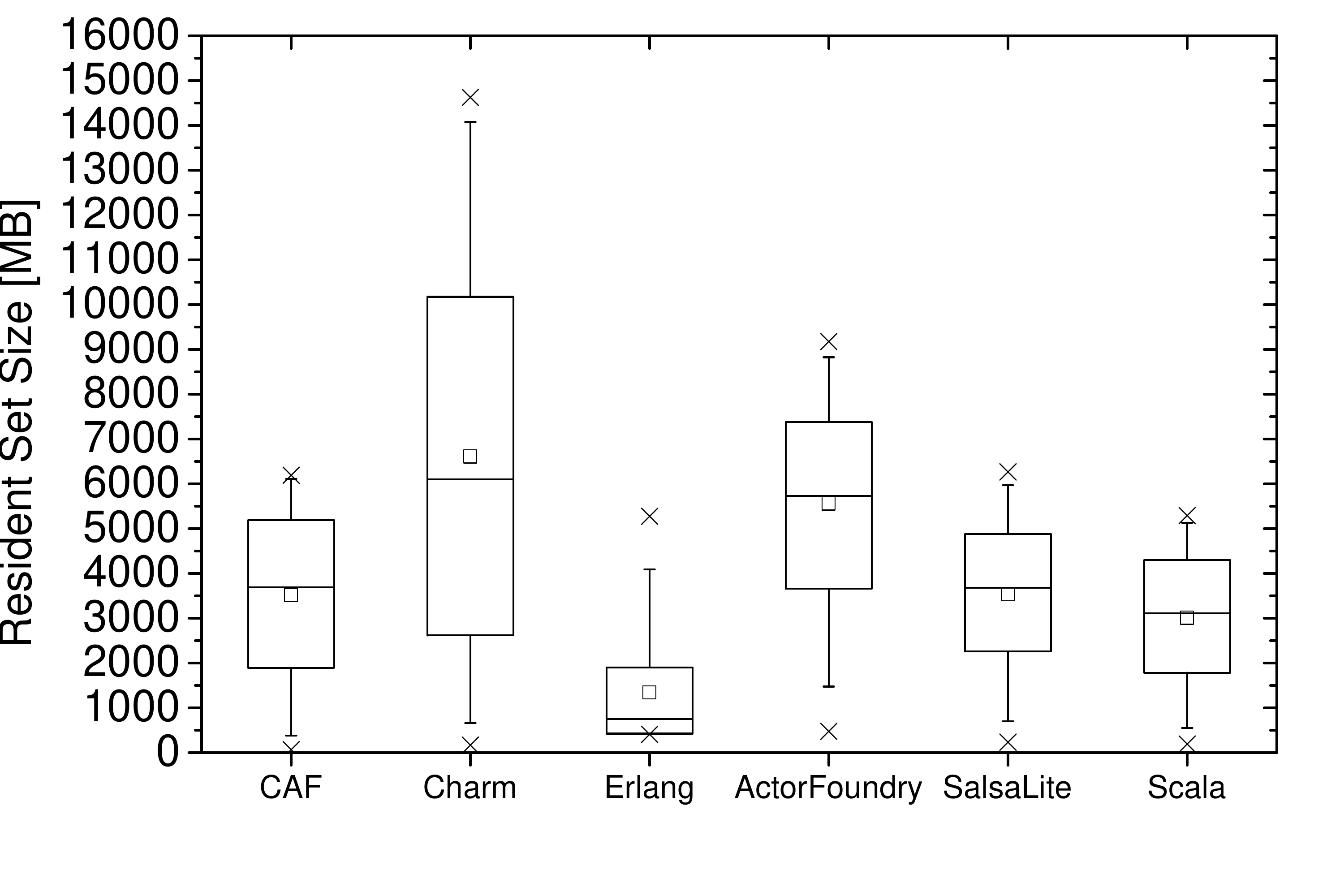}%
          {fig:mailbox:memory}%
          {Mailbox performance in N:1 communication scenario}%
          {fig:mailbox}

Figure~\ref{fig:mailbox:time} visualizes the time consumed by the applications to send and process the 100,000,000 messages as a function of available CPU cores.
As expected, all actor implementations show a steady growths of runtime on average, but differ significantly in values and fluctuations.
As an extreme, the performance of Erlang jumps by about an order of magnitude indicating a largely discontinuous resource scheduling.
The overall slopes differ greatly.
While CAF has a slope of 0.4, SALSA Lite is at 1.64 and Charm++ at 2.97.
The others rise even faster, limited by Erlang at 23.6.
However, the tail slopes above 32 cores can be separated into two groups.
While the first one with CAF, Charm and Salsa presents good scalability with a slope around 0, the remaining three implementations do not scale as well and have slopes between 6.2 and 6.7.
CAF outperforms all competitors in absolute values, underlining its strong level of optimization. 
On 64 cores, CAF has an average runtime of 104 seconds, which is about a tenth of the 1086 seconds measured for Scala.

Figure~\ref{fig:mailbox:memory} shows the resident set size during the benchmark execution.
In this scenario, a low memory usage can hint to a performance bottleneck, as 100 writers should be able to fill a mailbox faster than one single reader can drain it.
Erlang seems to deliver a good trade-off between runtime and memory consumption at first, but fails to maintain a reasonable runtime for high levels of hardware concurrency.
All three JVM-hosted application have a high memory consumption while running significantly slower than the native programs on average, indicating that writers do block readers and messages accumulate in the mailbox while the receiver is unable to dequeue them due to synchronization issues.
Compared to the other memory plots, this depicts a high spread for all implementations.
Over the course of the benchmark, the mailbox of the receiver fills with messages until all senders are done and no new messages arrive. At that point it is emptied again.
Hence, the plot summarizes the mailbox growth over this time.

\subsection{Mixed Operations Under Work Load}

In this benchmark, we consider a realistic use case including a mixture of operations under heavy work load.
The benchmark program creates a simple multi-ring topology with a fixed number of actors per ring.
A token with an initial value of 1,000 is passed along the ring and its value is decremented by one in each round.
A client that receives the token forwards it to its neighbor and terminates whenever the value of the token is 0.
Each of the 100 rings consists of 100 actors and is re-created 4 times.
Thus, we continuously create and terminate actors with a constant stream of messages.
In addition, one worker per ring performs prime factorization to add numeric work load to the system.

\doublefig{Sending and processing time}%
          {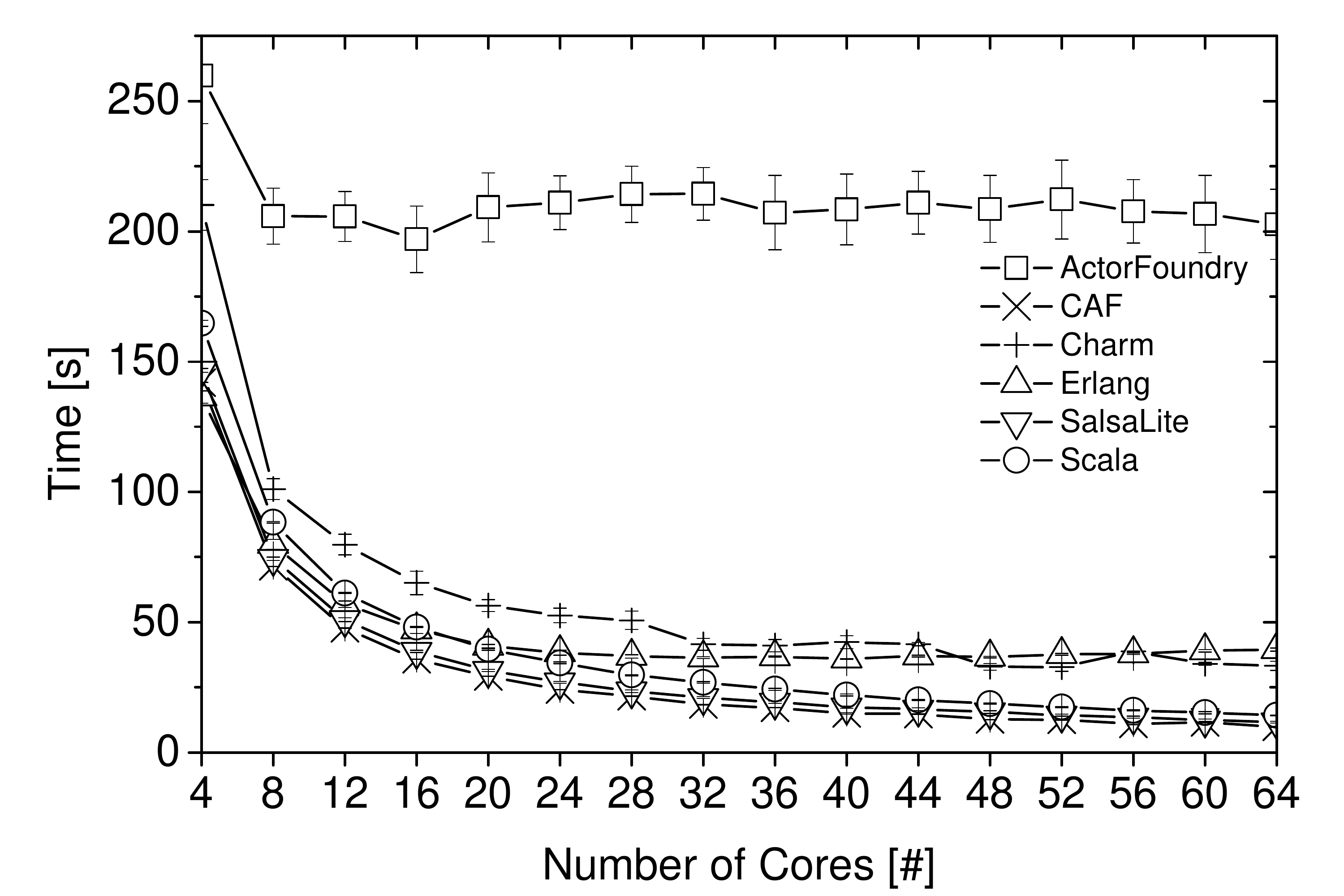}%
          {fig:mixed:time}%
          {Memory consumption}%
          {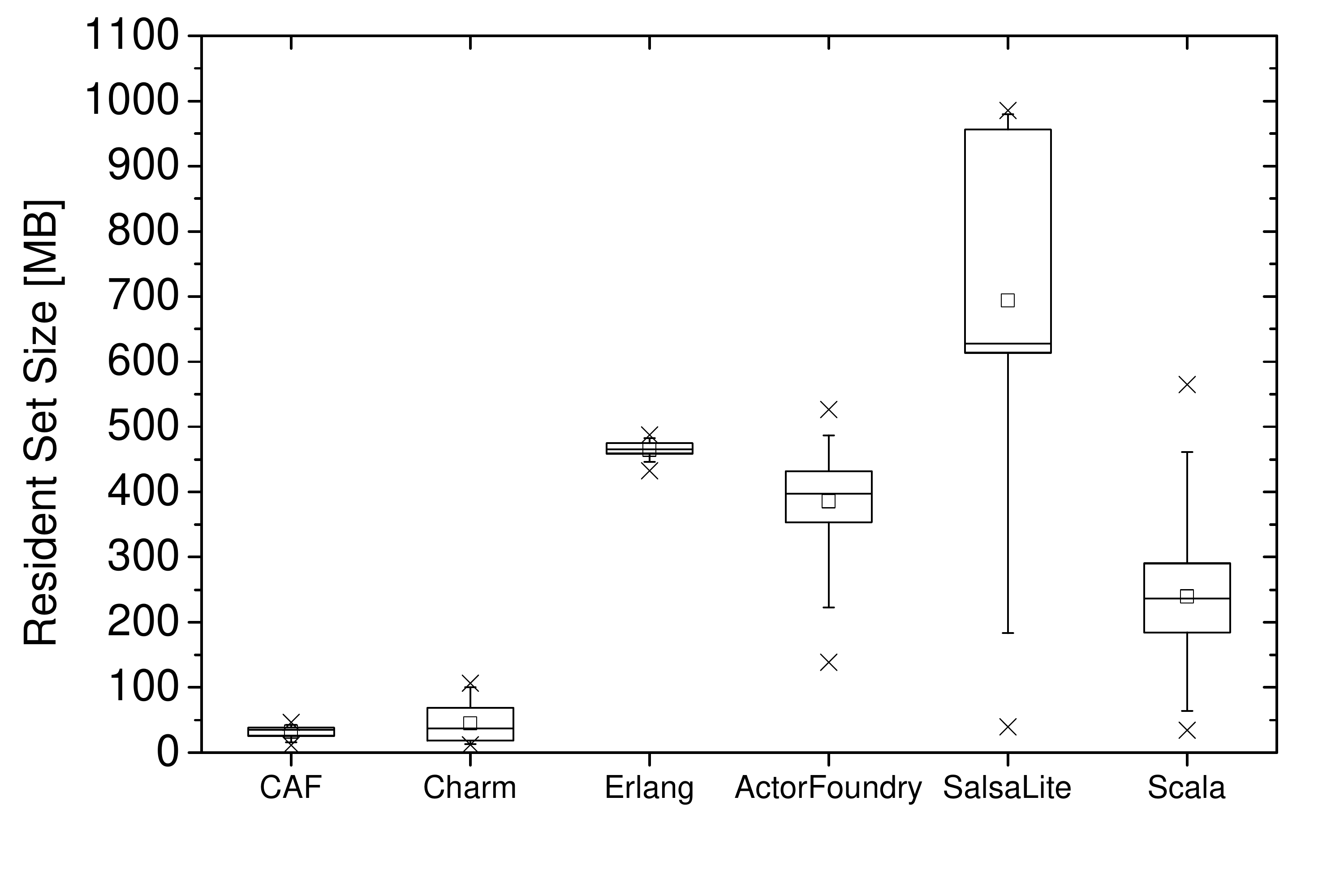}%
          {fig:mixed:memory}%
          {Performance in a mixed scenario with additional work load}%
          {fig:mixed}

Figure~\ref{fig:mixed:time} shows the runtime behavior as a function of available CPU cores.
Ideal scaling would halve the runtime when the number of cores doubles.
ActorFoundry stands out as it remains at a runtime above 200 seconds.
The process viewer from the operating system revealed that the benchmark program for ActorFoundry never utilizes more than five CPU cores at a time, why a better scalability cannot be expected.
SALSA Lite is the only implementation under test that performs similar to CAF in this benchmark, followed by Akka which is slightly slower.
It is worth mentioning that SALSA Lite required a manual work load distribution by the programmer.
Without putting each ring into its own ``Stage''---a scheduling unit in SALSA Lite---, the runtime increases by a factor between 10 and 20.

Figure~\ref{fig:mixed:memory} shows the memory consumption during the mixed scenario. Qualitatively, these values coincide well with our first benchmark results on actor creation.
CAF again has a very constant and thus predictable memory footprint, while using significantly less memory than all other implementations (below 50\,MB).
Noticeably, SALSA Lite, Erlang and ActorFoundry allocate more than factor ten of memory allocates by CAF.
However, SALSA uses the most memory of all benchmark programs and shows good performances, indicating that it may trade memory consumption for runtime efficiency.

\singlefig{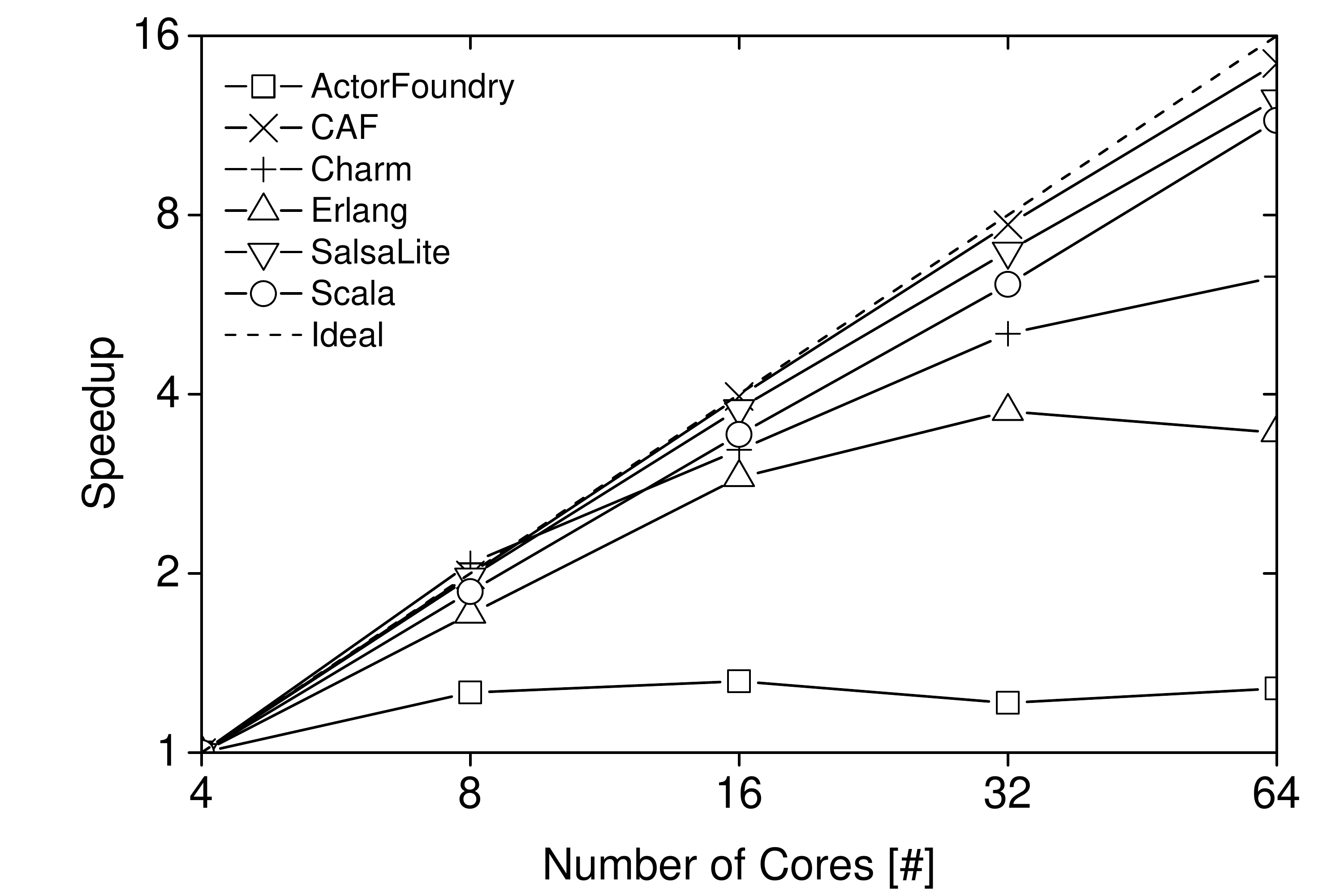}{Scaling behavior of the mixed case benchmark compared to an ideal speedup.}{fig:mixed:normalized}

Visualizing the runtime leads to similar curves for all implementations but the ActorFoundry.
To further examine the scaling behavior, we normalized each curve in relation to its performance on 4 cores.
The results are shown in Figure~\ref{fig:mixed:normalized} which displays the speedup of the mixed case benchmark as a function of the available cores using logarithmic scaling on both axes.
An ideal linear speedup, indicating that doubling the number of cores leads to twice the performance, is plotted as a dashed line.
Benchmarks that are close to the ideal line or have a similar slope exhibit scaling behavior.
Overall, CAF is closest to the ideal line, followed by SALSA and Scala.
Charm++ and Erlang show good performance up to 16 cores, but have decreased speedup thereafter.
Lastly, and consistent with Figure~\ref{fig:mixed:time}, ActorFoundry has the largest distance to the ideal curve indicating that it does not scale well in this benchmark.
It is noteworthy that all curves have an increasing deviation from the ideal behavior as the number of cores increases.
In relative values the speedup increase of CAF is in a range from 99\% to 93\% for each doubling in cores.
While SALSA is slightly behind decreasing to 90\%, Scala fluctuates close to 93\%.
Erlang exhibits a rapid decrease in speedup, starting around 100\% and rapidly falling towards 62\% as can be seen in its negative slope after 32 cores.
On 8 cores, Charm++ appears to have a better than linear scaling which is explained by high variances in its measurements.
Thereafter, its speedup decreases, mimicking the progression Erlang presented up to 32 cores.

\subsection{Computer Language Benchmarks Game}

\singlefig{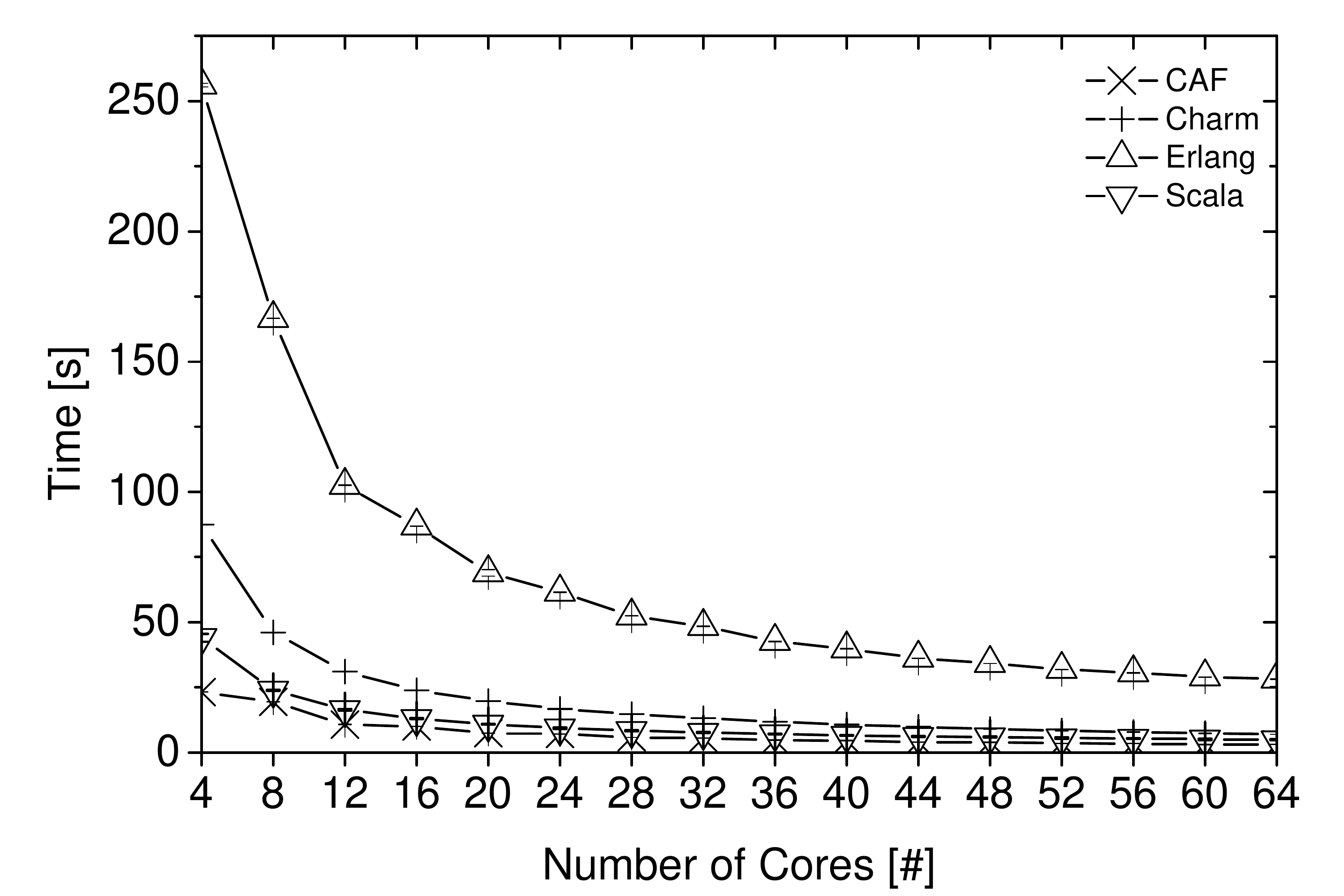}{Performance for calculating a Mandelbrot, adapted from the Computer Language Benchmarks Game}{fig:mandelbrot:runtime}

The Computer Language Benchmarks Game is a publicly available benchmark collection hosted by the Debian community\footnote{http://benchmarksgame.alioth.debian.org}.
It compares implementations of specific problems in different programming languages.
These benchmarks were written by community members and provide a standard way to compare different implementations of the same algorithms.

Among others, the benchmarks game offers the calculation of the Mandelbrot set, which we chose for our evaluation.
The calculation of the Mandelbrot set is a straight forward algorithm that parallelizes at fine granularity. 
The benchmark plots an $N$-by-$N$ pixel Mandelbrot set in the area $[-1.5-i,0.5+i]$.
While the original benchmark writes the resulting bitmap to a file, we chose to omit the output as we are not interested in I/O performance.
Each program distributes the problem by creating one actor per calculated row.
In contrast to the publicly available results, we plotted results from 4 to 64 cores in steps of 4, consistent with our previous experiments.
We consistently use a problem size of $N=16000$ and increased the iteration maximum from 50 to 500.
This increase provides us with a problem that is complex enough to observe scaling behavior up to 64 cores.

Our benchmark implementations are modified versions of the x64 Ubuntu quad-core programs.
We ported the available code from threads to the actor frameworks under test were needed.
Even if other solutions might be faster, they could not offer the features provided by the actor model---as considered in this paper.
The Erlang implementation is the unchanged version from the website and uses HiPE.
For Scala, we chose the unnumbered Scala benchmark and adapted it to use Akka actors.
The CAF benchmark is adapted from the C++ benchmark \#9 and uses CAF for parallelization instead of OpenMP.
As Charm++ is also based on C++, it uses the identical implementation for the Mandelbrot set.
However, parallelization in Charm++ did not work as expected.
We observed a drop in runtime after separating actor creation and message passing into two loops instead of one.
This is surprising, since both versions finished the loops nearly instantly, but afterwords required different times for the remaining calculations.
Furthermore, a straight forward implementation in a way similar to our other Charm++ benchmarks did not distribute the workload over all cores.
We improved the performance of Charm++ by assigning an equal fraction of actors to all cores dynamically at runtime, which reduced the runtime significantly. Due to the previous slow results, we excluded ActorFoundry from this competition even though Java versions of the algorithm exist.
Further, we do not have measurements for SALSA Lite as an implementation in its language was not available.

Figure~\ref{fig:mandelbrot:runtime} shows the runtime as a function of the available CPU cores.
Even though all benchmarks show a good overall scalability, their runtime varies largely.
CAF shows the best performance in this benchmark with a runtime of 3.2 seconds on 64 cores, followed by Scala at around 4.9 seconds.
Charm++ requires 7.0 seconds and Erlang performs worst at 28.2 seconds, which is more than CAF requires on 4 cores.
Since the benchmark focuses on number crunching, the performance of Erlang does not surprise as the VM of Erlang does not perform competitively for heavy numeric calculations.
However, we were surprised by the performance difference between Charm++ and CAF.
Even though both use the identical code for calculating the Mandelbrot set and performed very similar on the actor creation benchmark, Charm++ requires twice the runtime.
Since both frameworks use a non-preemptive scheduler, the performance difference must be the result of overhead in the runtime environment. 
We do not display memory measurements for this benchmarking task, as results plainly reflect the size of the pixel array for the Mandelbrot image set.

\subsection{Integrating OpenCL Actors}

\doublefig{Large problem: Mandelbrot set with 1.200 maximum iterations}
          {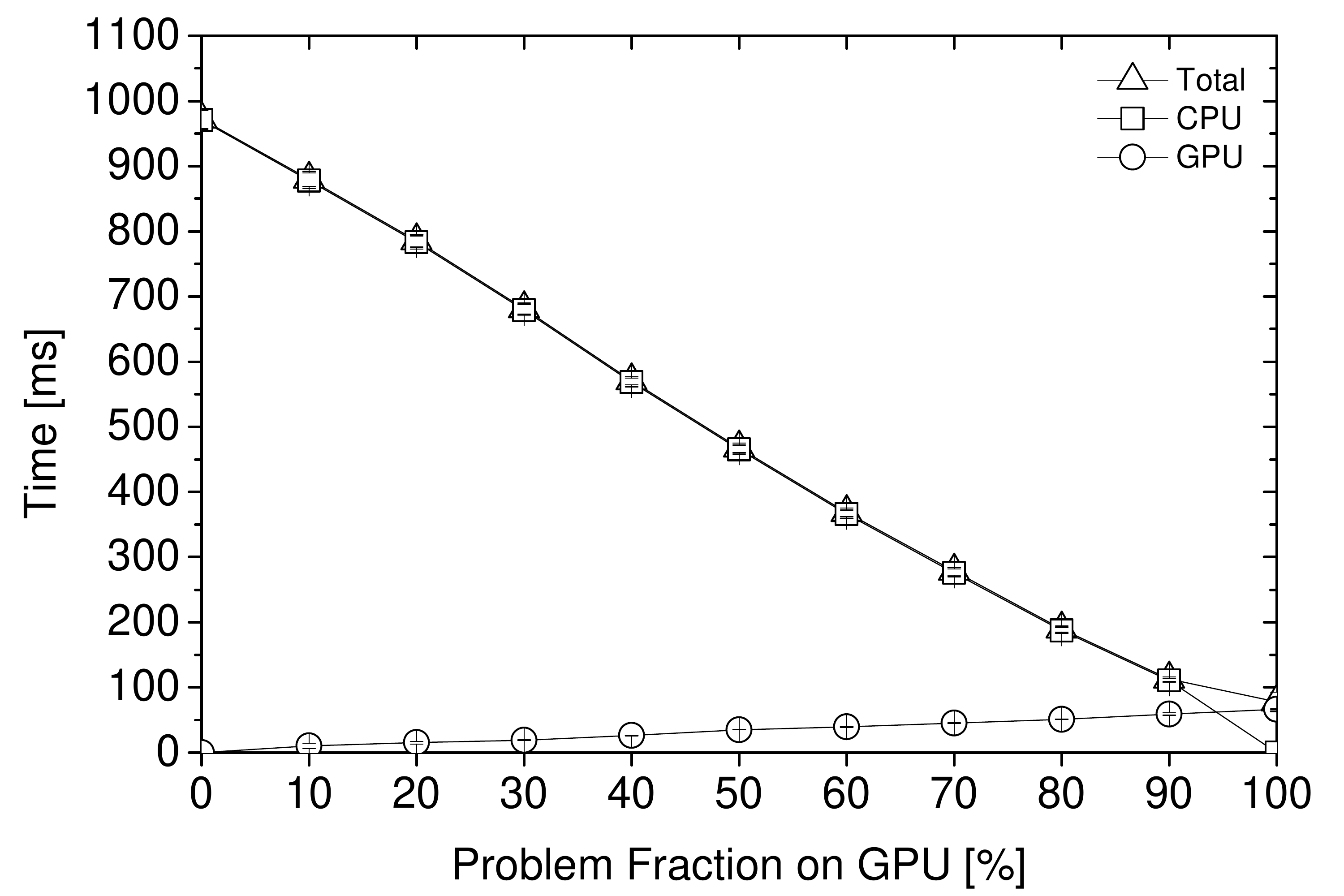}
          {fig:opencl:big}
          {Small problem: Mandelbrot set with 120 maximum iterations}
          {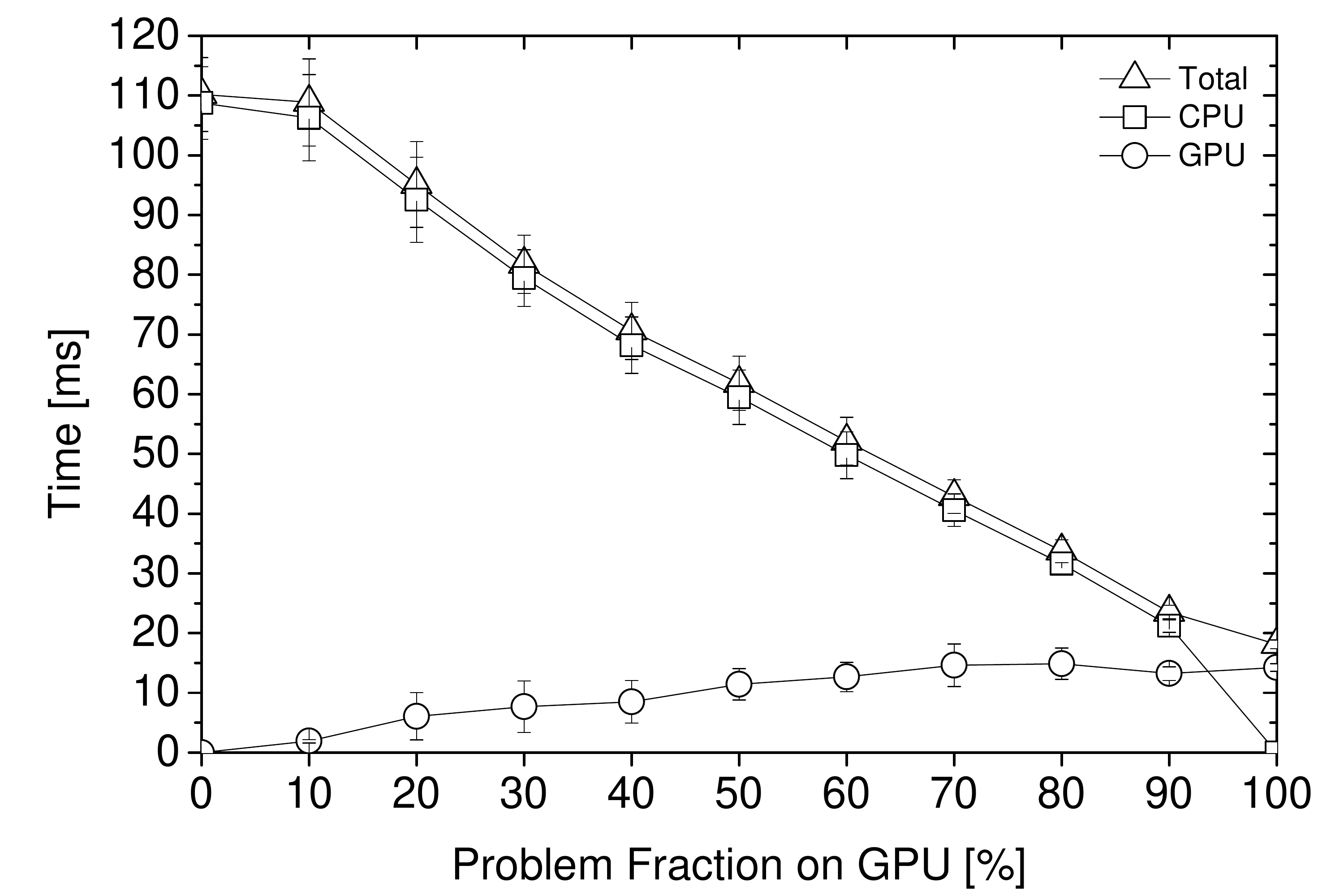}
          {fig:opencl:small}
          {Moving workload to the GPU via OpenCL actors.}
          {fig:opencl}

It is well known that GPUs vastly outperform CPUs for suitable problems. Such problems need to be divisible for concurrently executing on these SIMD machines. The Mandelbrot set used in the Computer Language Benchmarks Game falls into this category and is easily ported to OpenCL. In this benchmark we focus on the scalability of our OpenCL actor interface, and want to detect the overheads from offloading parts of our problem to the GPU.

This benchmark was performed on a machine equipped with two hexacore Intel Xeon CPUs clocked at 2.4 GHz and a Tesla C2075 GPU. It runs Linux and uses the OpenCL drivers provided by Nvidia. The  workload of our experiment is a cut from the inner part of a Mandelbrot set that has a balanced processing complexity for the entire image. We measured the wall-clock time for 11 different problem distributions, stepwise pushing a linearly increasing part of the calculation to the GPU. Starting from 0\,\% (full problem on CPU), we offloaded in steps of 10\,\% up to 100\,\% of the workload to the GPU. Furthermore, we measured two different problem sizes to examine the difference in scaling. Our workload is an image with a resolution of $1920x1080$ pixels in the area $[-0.5-0.7375i,0.1-0.1375i]$ with a different number of maximum iterations. For each measurement we performed 100 runs and plotted the mean as well as error bars that show the standard deviation.

Both graphs in Figure~\ref{fig:opencl} depict the  runtimes measured as  functions of the problem fraction offloaded to the GPU. Though the runtime is measured in milliseconds in both cases, the scales differ by an order of magnitude according to the different problem sizes. The problem evaluated in Figure~\ref{fig:opencl:big} is ten times larger than in Figure~\ref{fig:opencl:small}. In addition to the total runtime, the graphs include the runtime for the CPU and GPU calculations only, i.e., the time between starting all actors and their termination. The total runtime is not a sum of both as the calculations are performed in parallel. Overall the CPU runtime is slightly lower than the total runtime whenever CPU is used for the calculation. In comparison, the GPU requires a fraction of the CPU for processing in all cases the CPU is included in the calculation.

Overall, Figure~\ref{fig:opencl:big} shows excellent scalability with a clean linear decay of CPU consumption---the runtime falls at appropriate rate until execution on the GPU becomes dominant ($> 90 \%$).  The error bars are small compared to the runtime and nearly invisible in the graph. In contrast, Figure~\ref{fig:opencl:small} presents visible overhead. Because  the runtime of the offloaded background tasks is short, the penalty from feeding the Tesla with  code and data degrades the speed-up of offloading. Transferring a sub-millisecond task to the GPU does indeed create little performance improvements. Larger tasks again lead to linear program acceleration, but an overhead penalty of a few milliseconds remains as the gap between total and CPU-bound execution times. Larger error bars indicate that major parts of these operations are performed (and scheduled) by the operating system. 
In both test series, it takes longer to calculate 10~\% of the problem on the CPU than is needed to calculate 100~\% on the GPU. As a result, the lower bound is the time required to process the complete workload on the GPU.

In summary, these experiments on a large Tesla GPGPU reveal excellent scalability of programming GPUs with CAF actors. Significant executions attain an ideal speed-up, while the limit of offloading efficiency was detected for sub-millisecond tasks. Experiments with (smaller) desktop GPUs showed a lower efficiency threshold. 

\subsection{Discussion}

The first part of our analysis focused on tasks performed as part of the runtime environment.
Even short tasks with a small overhead can have a large impact on the runtime of an application.
Specifically, we performed two micro benchmarks to examine actor creation and the reception of massive amounts of messages.
Real world applications depend on the interaction of all tasks performed by the runtime in addition to their own logic.
In the extreme, previously well performing task may block each other or compete for resources when combined, or compete for resources with the applications logic.
The second half of our benchmarks focused on more balanced scenarios that efficiently make use of actors to solve a sample problem.
To increase the transparency of our testing, we included a community benchmark in this category.

In Section \ref{sec:case_for_actors}, we had started our discussion of  `actors in the wild' with the use case {\em elastic programming for adaptive deployment}. It essentially implies the demands for very low overhead from actor creation and high scalability of the overall runtime system. Our extensive evaluations target at these key indicators and reveal a diverse picture. Only SALSA Lite and CAF scale clearly at a very low memory footprint in actor creation, even though SALSA Lite wastes memory in the mixed use case. Partially, Scala and Charm++ also show proper performance, even though Charm++ flaws for high concurrency, and actor creation in Scala is 50 times less lean than in CAF. 

CAF and  Charm++---the only C++ competitors---remain apart. CAF runs faster, scales better, and uses less memory. 
It should be mentioned, though, that Charm++ is optimized for performance on clusters and supercomputers and as a direct result may not be as efficient at single-host performance.
Still, there are overlapping use cases for those systems that make a comparison justifiable.
For our runtime comparison, we have used the standalone version of Charm++ instead of its \texttt{charmrun} launcher that can be used to distribute an application or parallelize it using processes.

On the overall, CAF consistently approximated ideal scaling up to 64 cores and required significantly less memory than the competitors.
SALSA Lite and Scala revealed similar performance characteristics in some scenarios, but no competitor could reach the memory efficiency of CAF.
The mailbox performance benchmark is the only case where CAF consumes more memory than Erlang and Scala.
However, the higher memory allocation is a direct result of its highly scalable, lock-free mailbox  that allows CAF to outperform competing implementation by about an order of magnitude on 64 cores.

We highlighted the {\em end-to-end messaging at loose coupling} as second paradigm  in Section \ref{sec:case_for_actors}. This design concept has particular relevance for the IoT, where constrained devices are common.  Because of its consistently low memory footprint, but also its low computational overhead, CAF can be deployed in such environments of strictly limited  resources.
As the IoT is  home to a wide range of distributed applications that rely on message passing, it is worth introducing the actor model as a high level abstraction for developing applications. Charm++ , the other native solution of reasonable memory consumption could be candidate for IoT programming, as well. However, no adaptations to this domain are currently visible.

{\em Seamless integration of heterogeneous hardware} was our final area in focus when discussing key motivations for the actor model in Section \ref{sec:case_for_actors}. Offloading work to a GPU via the OpenCL actor offers a scalable solution to process heavy calculations at a low cost. While this can improve performance for small problems (from $1 ms$\ onwards), the gain in performance scales up ideally for large workloads.

\section{Conclusions and Outlook}
\label{sec:conclusion}

Currently the community faces the need for software environments that provide high scalability, robustness, and adaptivity to concurrent as well as widely distributed regimes.
In various use cases, the actor model has been recognized as an excellent paradigmatic fundament for such systems. Still, there is  a lack of full-fledged programming frameworks, which  in particular holds for the native domain. 

In this paper, we presented CAF, the ``C++ Actor Framework''.
CAF scales up to millions of actors on many dozens of processors including GPGPUs, and down to small systems---like Rasberry PIs \cite{hcs-eatdp-14}---in loosely coupled environments as are characteristic for the IoT.
We introduced a series of concepts to advance the reality of actor programming, most notably (a) a scalable, message-transparent architecture, (b) type-safe messaging interfaces between loosely coupled components with pattern matching, (c) an advanced scheduling facility. Together with ultra-fast, lean algorithms in the CAF core, this gave rise to consistently strong benchmark results of CAF, which clearly confirmed its excellent performance.

There are five future research directions.
Currently, we are reducing the resource footprint of CAF even further and port to the micro-kernel IoT operating system RIOT \cite{bhgws-rotoi-13}.
Second we work on extending scheduling and load sharing to distributed deployment cases and massively parallel systems. This work will stimulate further, compatible benchmarking \cite{is-sabs-14}.
Third we will extend our design towards  effective monitoring and debugging facilities.
Fourth we explore design space opened up by the typed messaging interfaces for composable actor handles.
Finally, a robust security layer is on our schedule that subsumes strong authentication of actors in combination with opportunistic encryption.

\section*{Acknowledgements}

The authors would like to thank Matthias Vallentin and Matthias W\"ahlisch, who had a helping hand or three in shaping CAF. Alex Mantel and Marian Triebe contributed to implementing benchmarks and runtime inspection programs, as well as testing and bugfixing.
We further want to thank the Hamburg iNET working group for vivid discussions and inspiring suggestions.
Funding by the German Federal Ministry of Education and Research within the projects ScaleCast, SAFEST, and Peeroskop is gratefully acknowledged.

\bibliography{}

\end{document}